\documentclass[12pt]{iopart}

\usepackage{graphicx}
\usepackage{iopams}

\begin{document}

\title{Qubit-nonlinear-oscillator systems: from the moderate-coupling limit to the ultrastrong-coupling regime}
\author{O de los Santos-S\'anchez} 

\address{Instituto de Ciencias F\'{\i}sicas, Universidad
Nacional Aut\'onoma de M\'exico, Apdo. Postal 48-3, 
Cuernavaca, Morelos 62251, M\'exico}

\eads{\mailto{octavio.desantos@gmail.com}}

\begin{abstract}
This work aims to provide an alternative approach to modeling a two-state system (qubit) coupled to a nonlinear oscillator. Within a single algebraic scheme based upon the f-deformed oscillator description, hard and soft nonlinearities are proposed to be simulated by making use of fitting algebraic models extracted from the trigonometric and modified P\"oschl-Teller potentials, respectively. In the regime where the strength of the coupling is considered to be moderate, this approach allows for an analytic, albeit approximate, diagonalization process of the proposed Hamiltonian through using the Van Vleck perturbation theory and embracing the two types of nonlinear features. In the ultrastrong-coupling regime, the effect of such nonlinearities upon the squeezing and phase space properties of the ground state of the composite system is also explored by numerical means. 
\end{abstract}


\maketitle

\section{Introduction}\label{sec:1}

A two-state system coupled to a harmonic oscillator epitomizes one of the most versatile, and ubiquitous, bipartite  models of interaction suitable to describe, accurately, a wide variety of context-dependent physical implementations in which the richness of its quantum predictions has been embodied in different coupling regimes. A number of archetypal examples of the foregoing for quantum computation, involving the use of either artificial or natural atoms generally modeled as two-state systems, can be quoted \cite{buluta}, scenarios where, for instance, a two-level quantum dot can be strongly coupled to a photonic crystal cavity \cite{faraon} or a semiconductor microcavity \cite{reith}; entanglement experiments involving a Rydberg two-level atom interacting with a cavity radiation field via its dipole moment \cite{haroche}; and realizations concerning a quantum biological description of light-harvesting complexes that can, in principle, be emulated with modern circuit quantum electrodynamics (QED) systems \cite{altintas}. In this last respect, interest in solid-state realizations of artificial qubit-oscillator systems has also been growing in recent years, pushed by the technological development of Cooper-pair boxes \cite{makhlin} coupled to superconducting (SC) transmission-line resonators \cite{blais,blais2,wallraff} and Josephson flux qubits \cite{orlando} read out by superconducting quantum interference devices (SQUID's) \cite{chiorescu,johanson}. Indeed, these new technologies have paved the way for  implementing hybrid systems (combining elements of atomic physics, molecular physics, and quantum optics, to mention some examples) incorporating SC qubits coupled to oscillator configurations such as waveguide resonators or SQUID's, admitting, in turn, the possibility of adjusting the parameter settings, on demand, and exploring new quantum phenomena \cite{xiang,xiangyou}.  \\

Current circuit QED architectures, in which a Josephson junction is embedded in a microwave transmission-line cavity, have also allowed, by adjusting the parameters of the circuit, engineering a wide range of Kerr-type nonlinearities in these implementations, opening a new pathway to realize, for example, on the basis of a nonlinear resonator description, the fast generation of Schr\"oringer cat states of microwave light \cite{bourassa}. And, among other areas of application, one can mention the implementation of bifurcation \cite{metcalfe} or parametric \cite{castellanos} amplifiers useful to read out the state of an artificial qubit, and the exploration of squeezing effects of microwave light \cite{castellanos}. Furthermore, it has also been experimentally demonstrated in \cite{gerhard} the generation of multi-component Shr\"odinger-cat-like states by engineering an artificial Kerr medium through 3D circuit QED architectures. In this context, it is also worth mentioning the interest in quantifying quantum fluctuations revealed by the emission spectrum of driven nonlinear oscillators \cite{andre}, the possibility of observing multiphoton blockade \cite{adam}, and the spectrum itself of a superconducting qubit dispersively coupled to a nonlinear oscillator \cite{ong}.  \\

So, in the light of the aforementioned studies, it makes sense to consider the nonlinear regime of qubit-oscillator configurations as a relevant research issue from the quantum theoretical point of view. Actually, the presence of such nonlinearities makes it difficult to make use of standard algebraic methods to get manageable analytic results when the interplay between the qubit and the nonlinear oscillator falls into the limit of the moderate- or ultrastrong-coupling regime (this means that the qubit-oscillator coupling is a significant fraction of the qubit's transition frequency, a regime  for which the rotating-wave approximation is known to fail), and within which systems of current interest, such as those previously mentioned in the context of circuit QED, predominantly operate. This work is mainly focused on putting forward an alternative, and generic, algebraic description of quantum qubit-nonlinear-oscillator systems encompassing two types of nonlinearities ({\it i.e.,} hard and soft nonlinearities $\propto \chi (\hat{a}^{\dagger}\hat{a})^{2}/2$ for which, to be more precise, the value of the anharmonic coefficient $\chi$ can be either positive or negative) within a single algebraic framework with the help of the f-deformed oscillator scheme \cite{manko}; besides the set of examples mentioned at the outset, the proposed model itself may additionally be suitable in describing molecular scenarios (polar molecules viewed as quantum bits) \cite{cusati,napoli,zoller} where the possibility of featuring anharmonic effects could also play an important descriptive role. So, to deal with this issue, in the weak nonlinear regime (that is to say, $\chi \ll 1$), we find it convenient to deploy P\"oschl-Teller oscillators to simulate, by properly choosing a deformation function, such nonlinear features and extracting from them a suitable approximate representation of the observables characterizing the position and momentum variables associated with each oscillator; this will be the theme of section \ref{sec:2} together with the content delineated in \ref{appen:a}. In section \ref{sec:3}, the qubit-nonlinear-oscillator model is introduced and its eigenvalues are obtained, in the moderate-coupling regime, with the aid of Van Vleck perturbation theory, a procedure inspired by the works of Hausinger and Grifoni \cite{grifoni1} and Vierheilig {\it et al.} \cite{grifoni2}; the latter being restricted to the case of hard nonlinearites in the context of a quartic oscillator. The limitations of such a procedure are outlined in section \ref{sec:4} by numerically diagonalizing an extended version of the proposed model, allowing us to further explore some properties of the ground state of the composite system, such as squeezing and phase space picture, within the first stages of the ultrastrong-coupling regime. Finally, some conclusions and perspectives are given in section \ref{sec:5}.

\section{Symmetric P\"oschl-Teller-like systems as nonlinear oscillator models}\label{sec:2}

According to Man'ko and co-workers \cite{manko}, a f-oscillator is a non-harmonic quantum system characterized by an algebraic Hamiltonian of the form 
\begin{equation}\label{eq:hdef1}
\hat{H}_{D} = \frac{\hbar \Omega}{2}\left(\hat{A}^{\dagger}\hat{A}+\hat{A}\hat{A}^{\dagger} \right),
\end{equation}
where $\Omega$ is a frequency parameter, and the generalized boson annihilation and creation operators, $\hat{A}$ and $\hat{A}^{\dagger}$, respectively, are constructed by deforming their harmonic counterparts, $\hat{a}$ and $\hat{a}^{\dagger}$, through the non-canonical transformation
\begin{equation}
\hat{A} = \hat{a}f(\hat{n}) = f(\hat{n}+1)\hat{a}, \qquad \hat{A}^{\dagger} = f(\hat{n})a^{\dagger} = \hat{a}^{\dagger} f(\hat{n}+1), \label{eq:demormedops}
\end{equation}
with $f(\hat{n})$ being a deformation function that takes on the role of capturing, at least partially at this stage, via the number operator $\hat{n}=\hat{a}^{\dagger}\hat{a}$, the level-excitation dependency of a given nonlinear system having a non-equidistant energy spectrum; and, the same function may also be dependent on any other (or more than one) fitting physical parameter. With the help of the standard commutation relation $[\hat{a},\hat{a}^{\dagger}]=1$, the aforesaid Hamiltonian can equivalently be cast in terms of the number operator as
\begin{equation}
\hat{H}_{D} = \frac{\hbar \Omega}{2} \left( (\hat{n}+1)f^{2}(\hat{n}+1)+\hat{n}f^{2}(\hat{n})  \right).
\label{eq:hamdef2}
\end{equation}
In this algebraic scheme, the deformed ladder operators (\ref{eq:demormedops}) have the following effect on the number operator basis
\begin{equation}
\hat{A}|n\rangle = f(n)\sqrt{n}|n-1\rangle, \qquad \hat{A}^{\dagger}|n\rangle = f(n+1)\sqrt{n+1}|n+1\rangle,
\label{eq:actiondef}
\end{equation}
and they also obey the commutation relationships  
\begin{equation} \fl 
[\hat{A}, \hat{n}] = \hat{A}, \qquad [\hat{A}^{\dagger}, \hat{n} ] = -\hat{A}^{\dagger}, \qquad [ \hat{A},\hat{A}^{\dagger}] = (\hat{n}+1)f^{2}(\hat{n}+1)-\hat{n}f^{2}(\hat{n}).
\end{equation}
It is worth underlying the fact that, in contrast to the harmonic oscillator case, a f-deformed oscillator Hamiltonian is no longer a linear function of the number operator since additional powers of it may play a significant role in describing the inherent nonlinear features of the system under study. In the same sense, we note that the commutator between the deformed ladder operators ceases to be a c-number, since it can become a function of the number operator whose complexity will  depend on the particular choice of the deformation function. When referring to (\ref{eq:actiondef}), we also see that the deformed operators change the number of quanta in $\pm 1$ and their corresponding matrix elements are modified via the deformation process;  they create and annihilate anharmonic quanta within the context of  nonlinear systems possessing non-equidistant energy spectra. This reasoning is reaffirmed by the commutation property 
\begin{equation}
[\hat{H}_{D},\hat{A}] = -\hbar \Omega_{\hat{n}} \hat{A}, \qquad [\hat{H}_{D},\hat{A}^{\dagger}]= \hbar \hat{A}^{\dagger} \Omega_{\hat{n}},
\end{equation}
where, as is easily inferred from the algebraic Hamiltonian given by (\ref{eq:hamdef2}), $\Omega_{\hat{n}} = \frac{\Omega}{2}\left( (\hat{n}+2)f^{2}(\hat{n}+2)-\hat{n}f^{2}(\hat{n}) \right)$ is construed as the operator that quantifies the frequency separation between adjacent energy levels. Notice that the harmonic-oscillator algebra is retrieved in the limit $f(\hat{n}) \to 1$. 

Incidentally, this algebraic framework has formerly been applied to cope with the problem of constructing approximate coherent states defined either as eigenstates of the deformed annihilation operator or by application of a displacement-like operator to the ground state of anharmonic systems, such as the Morse \cite{recamier1,octa1,octa2} and symmetric P\"oschl-Teller potentials \cite{octa3}. The latter ones are the class of systems that will be of interest to us and, at this point, we briefly digress to introduce  them within their corresponding physical space.\\

So, depending on the physical scenario we are interested in, let the nonlinear system under consideration be described by any of the following two symmetric P\"oschl-Teller (PT) models: (i) The trigonometric potential (TPT) or (ii) the  hyperbolic one, also called modified potential (MPT), the former supporting an infinite number of bound states and the latter, meanwhile, supporting a finite number of bound states. The shape of the one-dimensional TPT potential is explicitly given by the expression
\begin{equation}
V_{+}(x) = U_{0}^{(+)}\tan^{2}(ax),
\end{equation}
where $U_{0}^{(+)}$ is the potential's strength and $a$ its range. By solving the Schr\"odinger equation for this confining system, the corresponding eigenfunctions and eigenenergies  are found to be given, respectively, by \cite{nieto1}
\begin{eqnarray}
\psi_{n}^{(\lambda_{+})}(x) & = &  \sqrt{\frac{a(\lambda_{+}+n)\Gamma(2\lambda_{+}+n)}{\Gamma(n+1)}} (\cos(ax))^{1/2} P_{n+\lambda_{+}-1/2}^{1/2-\lambda_{+}}(\sin(ax)), \label{eq:wavef1} \\
E_{n}^{(+)} & = & \hbar \Omega_{+} \left( n+\frac{1}{2}+\frac{n^{2}}{2\lambda_{+}} \right), \quad n=0,1,2 \ldots \label{eq:spectrum1}
\end{eqnarray}
Here, the $P_{j}^{i}$'s stand for the associated Legendre polynomials, $\Omega_{+}=\hbar a^{2}\lambda_{+}/\mu_{+}$, with $\mu_{+}$ being the effective mass of the oscillator system, and $\lambda_{+}$ is related to the potentials'  strength through the relation $\lambda_{+}(\lambda_{+}+1)=2\mu U_{0}^{(+)}/\hbar^{2}a^{2}$.\\

On the other hand, the MPT potential is represented by the hyperbolic expression 
\begin{equation}
V_{-}(x) = U_{0}^{(-)}\tanh^{2}(ax),
\end{equation}
where $U_{0}^{(-)}$ is the depth of the finite well and $a$ is the scale parameter. The wave functions associated with the bound part of its spectrum and the corresponding eigenenergies are \cite{nieto2}
\begin{eqnarray}
\psi_{n}^{(\lambda_{-})}(x) & = & \sqrt{\frac{a(\lambda_{-}-n)\Gamma(2\lambda_{-}-n+1)}{\Gamma(n+1)}} P_{\lambda_{-}}^{n-\lambda_{-}}(\tanh (a x)), \label{eq:wavef2}\\
E_{n}^{(-)} & = & \hbar \Omega_{-} \left( n+\frac{1}{2}-\frac{n^{2}}{2\lambda_{-}} \right), \quad n=1,2,\ldots ,n_{max}, \label{eq:spectrum2}
\end{eqnarray}
where, in the same sense as in the previous case, $\Omega_{-}=\hbar a^{2}\lambda_{-}/\mu_{-}$, save for the fact that $\lambda_{-}$ is now taken to be an integer related to the number of bound states of the system such that the last bound state corresponds to $n_{max}=\lambda_{-}-1$. Unless otherwise specified, throughout  the remainder of this work the subscripts (or superscripts) $\pm$ are chosen to label expressions (or quantities) belonging to TPT- and MPT-type systems, respectively. 

It is also noted in passing that, for sufficiently small values of $\lambda_{\pm}^{-1}$, but strong enough to be conspicuous, the shape of these outlined potentials closely resembles the undriven Duffing oscillator \cite{andre} (a quartic oscillator) involving hard or soft nonlinearities, $\pm \lambda_{\pm}$, as viewed from the former ones' series expansion, {\it i.e.}, $V_{\pm}(x) \approx \frac{1}{2}\Omega_{\pm}^{2}x^{2}\pm \frac{1}{3} \frac{\Omega^{3}_{\pm}}{\lambda_{\pm}}x^{4}+O(\lambda_{\pm}^{-2},x^{6})$; this is so provided that the oscillations are also restricted to the low-lying region of the potentials'  well. For the sake of emphasis, units such that $\hbar=\mu_{\pm}=1$ were taken to illustrate this particular fact, so that the quantity $\lambda^{-1}_{\pm}$ is chosen to take on the role of describing the degree of anharmonicity of the system in question. Thus, from now on, the symbolic parameter $\chi$ previously mentioned in the introduction shall be replaced by $\lambda_{\pm}^{-1}$. \\

So, returning to the subject of f-oscillators, and in the light of the foregoing, let the deformation function be chosen to have the following form \cite{octa3}
\begin{equation}
f^{2}_{\pm}(\hat{n}) = 1\pm \frac{\hat{n}-1}{2\lambda_{\pm}},
\label{eq:fdeformation}
\end{equation}
where each of the dimensionless parameters $\lambda_{\pm}$ is taken to characterize the degree of anharmonicity of the respective nonlinear oscillator. Substitution of this function into (\ref{eq:hamdef2}) leads to the concise Hamiltonian  
\begin{equation}
H_{D}^{(\pm)}  =  \hbar \Omega_{\pm} \left (\hat{n}+\frac{1}{2}\pm\frac{\hat{n}^{2}}{2\lambda_{\pm}} \right), \label{eq:spectrum3}
\end{equation}
whose $n$-dependent quadratic spectrum epitomizes that of the P\"oschl-Teller potentials outlined above (see  (\ref{eq:spectrum1}) and (\ref{eq:spectrum2})), with the identification $\Omega_{\pm} = \hbar a^{2}\lambda_{\pm}/\mu_{\pm}$. So, having established the deformation function, the action of the deformed operators upon the number state basis is also straightforwardly stated, namely,
\begin{equation} \fl 
\hat{A}_{\pm}|n\rangle =  \sqrt{n\left(1\pm \frac{n-1}{2\lambda_{\pm}} \right)} |n-1\rangle, \qquad
 \hat{A}^{\dagger}_{\pm}|n\rangle  =  \sqrt{(n+1)\left(1\pm \frac{n}{2\lambda_{\pm}} \right)} |n+1\rangle, \label{eq:fladder}
\end{equation}
along with the commutation relation 
\begin{equation}
[\hat{A}_{\pm}, \hat{A}^{\dagger}_{\pm} ] = 1\pm \frac{\hat{n}}{\lambda_{\pm}}. \label{eq:fcommute}
\end{equation}
Thus, the frequency separation between adjacent energy levels of the respective oscillator is found to be assessed by the opeartor 
\begin{equation}
\Omega_{\hat{n}}^{(\pm)} = \Omega_{\pm} \left ( 1\pm \frac{2\hat{n}+1}{2\lambda_{\pm}}\right ).
\label{eq:frecuencyn}
\end{equation}
We remark that all of the f-deformed algebra above contracts to the harmonic oscillator case ($f(\hat{n}) \to 1$) by taking the limit $\lambda_{\pm} \to \infty$ while keeping the product $a^{2}\lambda_{\pm} = \mu_{\pm} \Omega_{\pm}/\hbar$ finite. \\

The f-deformed oscillator turns out to be a particularly convenient algebraic tool because not only does it encode,  in a simple way, the nonlinear character of the system it seeks to describe for a given deformation function, but it also permits us to establish a connection between the deformed operators and the actual energy-raising and -lowering  operators obtained by means of standard factorization methods. For instance, from an algebraic-structure point of view, it transpires that there is a one-to-one correspondence between the actual jump operators for the independent cases of the Morse and the TPT potentials and the ones corresponding to the f-deformed-oscillator version of such systems in terms of their effective action upon the  eigenstates of the system at hand, including, of course, the equivalence between their commutation relationships \cite{octa1,ancheyta}. To be more explicit in this regard, as for the actual lowering and raising operators associated with the symmetric TPT potential, one gets \cite{dong}
\begin{eqnarray} \fl 
\hat{b}_{+} \psi_{n}^{(\lambda_{+})}(u)  =  \sqrt{n \left (1+\frac{n-1}{2\lambda_{+}} \right)} \psi_{n-1}^{(\lambda_{+})}(u), \  \hat{b}^{\dagger}_{+} \psi_{n}^{(\lambda_{+})}(u)  = \sqrt{(n+1)\left (1+\frac{n}{2\lambda_{+}} \right)} \psi_{n+1}^{(\lambda_{+})}(u), \label{eq:actionb}
\end{eqnarray}
where $\psi_{j}^{(\lambda_{+})}(u)$ are the TPT wave functions (\ref{eq:wavef1}), $u=\sin (ax)$, and the operators $\hat{b}_{+}$ and $\hat{b}^{\dagger}_{+}$, which turn out to obey the commutation relation
\begin{eqnarray}
[\hat{b}_{+}, \hat{b}^{\dagger}_{+}] = 1+\frac{\hat{n}_{+}}{\lambda_{+}}, \label{eq:commuteb}
\end{eqnarray}
providing that the number operator $\hat{n}_{+}$ is such that $\hat{n}_{+} \psi_{n}^{(\lambda_{+})} (u) = n\psi_{n}^{(\lambda_{+})}(u)$, have the differential form 
\begin{equation} \fl 
\hat{b}_{+} = (1-u^{2})\left(\frac{d}{du}+\frac{u\epsilon_{+}}{1-u^{2}} \right)\sqrt{\frac{\epsilon_{+}-1}{2\lambda_{+} \epsilon_{+}}}, \  \hat{b}^{\dagger}_{+}  =  (1-u^{2})\left(-\frac{d}{du}+\frac{u\epsilon_{+}}{1-u^{2}} \right)\sqrt{\frac{\epsilon_{+}+1}{2 \lambda_{+} \epsilon_{+}}}, \label{eq:diferential1}
\end{equation}
with $\epsilon_{+}=\lambda_{+}+n$. 

On the other hand, the description of the symmetric MPT oscillator is given in terms of the variable $u=\tanh (a x)$ as seen from (\ref{eq:wavef2}). Accordingly, with the help of the factorization method, it is possible to determine the action of the corresponding raising and lowering operators upon the MPT wave functions and establish their differential form. In doing so, one gets \cite{dong}
\begin{eqnarray}\fl 
\hat{b}_{-} \psi_{n}^{(\lambda_{-})}(u)  =  \sqrt{n\left(1-\frac{n-1}{2\lambda_{-}}\right)} \psi_{n-1}^{(\lambda_{-})}(u), \  \hat{b}^{\dagger}_{-} \psi_{n}^{(\lambda_{-})}(u)  = \sqrt{(n+1)\left(1-\frac{n}{2\lambda_{-}} \right)} \psi_{n+1}^{(\lambda_{-})}(u), \label{eq:actionbm}
\end{eqnarray}
where 
\begin{equation} \fl 
\hat{b}_{-} = \sqrt{1-u^{2}}\left(\frac{d}{du}+\frac{u\epsilon_{-}}{1-u^{2}} \right)\sqrt{\frac{\epsilon_{-}+1}{2\lambda_{-} \epsilon_{-}}}, \  \hat{b}^{\dagger}_{-}  =  \sqrt{1-u^{2}}\left(-\frac{d}{du}+\frac{u\epsilon_{-}}{1-u^{2}} \right)\sqrt{\frac{\epsilon_{-}-1}{2 \lambda_{-}\epsilon_{-}}}, \label{eq:diferential2}
\end{equation}
with $\epsilon_{-}=\lambda_{-}-n$. So, in this case, the operators $\hat{b}_{-}$ and $\hat{b}^{\dagger}_{-}$ satisfy  the commutator
\begin{eqnarray}
[\hat{b}_{-}, \hat{b}^{\dagger}_{-} ] = 1-\frac{\hat{n}_{-}}{\lambda_{-}}, \label{eq:commutebm}
\end{eqnarray}
with $\hat{n}_{-}$ being the corresponding number operator adopted in the same sense as in the previous case. \\

Thus, comparison of (\ref{eq:fladder}) and (\ref{eq:fcommute}) with (\ref{eq:actionb}), (\ref{eq:commuteb}) and (\ref{eq:actionbm}), (\ref{eq:commutebm}), respectively, allows us to realize better the algebraic equivalence between both techniques; this fact legitimates the use of the f-deformed oscillator as an algebraic scheme for encapsulating the kind of nonlinear features that such systems are endowed with. So, translating these features into a single framework (the f-deformed oscillator) also motivates us to strive to build an approximate representation of physical variables in terms of deformed operators able to describe, at least approximately, those anharmonic oscillations  inherent in the study of this class of systems. One way of dealing with this problem would be to establish a direct relation between the actual matrix elements of a given observable and the ones obtained by means of a series expansion of the same variable in terms of the deformed operators, a procedure that was originally proposed in \cite{carvajal} to describe vibrational excitations of molecules represented by Morse oscillators, in the framework of the su(2) one-dimensional version of the vibron model. However, since the matrix elements for the physical variables of interest to us, say, the coordinate and/or momentum, are not known in a simple closed form in the context of P\"oschl-Teller potentials, let us adopt an alternative strategy outlined in \cite{lemus}, which was illustrated for the MPT potential solely, which consists in inverting the differential form of the ladder operators to be able to extract from it either the coordinate $x$ or the momentum $\hat{p}$ variable; here, the aforesaid technique is applied to the TPT potential, while the MPT case is also incorporated for the sake of completeness and by virtue of the aim of the present work. Thus, in the first instance, allowing for the fitting and justified algebraic identifications $\hat{b}_{\pm} \to \hat{A}_{\pm}$ ($\hat{b}^{\dagger}_{\pm} \to \hat{A}^{\dagger}_{\pm}$), and $\hat{n}_{\pm}\to \hat{n}$, one can deduce from (\ref{eq:diferential1}) the following relation concerning the TPT potential
\begin{eqnarray}
\sin(ax_{+})  =  \frac{1}{\sqrt{2\lambda_{+}}} \left (\hat{A}^{\dagger}_{+}F_{\hat{n}}^{(+)}+\hat{A}_{+} G_{\hat{n}}^{(+)} \right),
\end{eqnarray}
where we have set the diagonal operators
\begin{displaymath}
F_{\hat{n}}^{(+)}  =  \frac{1}{\sqrt{\left (1+\frac{\hat{n}}{\lambda_{+}}\right )\left (1+\frac{\hat{n}+1}{\lambda_{+}}\right)}}, \qquad G_{\hat{n}}^{(+)} =  \frac{1}{\sqrt{\left (1+\frac{\hat{n}}{\lambda_{+}}\right )\left(1+\frac{\hat{n}-1}{\lambda_{+}}\right)}}.
\end{displaymath}
In like maner, from (\ref{eq:diferential2}), one gets for the MPT potential \cite{lemus}
\begin{eqnarray}
\sinh(ax_{-})  =  \frac{1}{\sqrt{2\lambda_{-}}} \left (\hat{A}^{\dagger}_{-}F_{\hat{n}}^{(-)}+\hat{A}_{-} G_{\hat{n}}^{(-)} \right),
\end{eqnarray}
with  
\begin{displaymath}
F_{\hat{n}}^{(-)}  =  \frac{1}{\sqrt{\left (1-\frac{\hat{n}}{\lambda_{-}}\right )\left (1-\frac{\hat{n}+1}{\lambda_{-}}\right )}}, \qquad G_{\hat{n}}^{(-)} =  \frac{1}{\sqrt{\left (1-\frac{\hat{n}}{\lambda_{-}}\right )\left (1-\frac{\hat{n}-1}{\lambda_{-}}\right)}}. 
\end{displaymath}
Successive approximations for the coordinate $x_{\pm}$ (or momentum $\hat{p}_{\pm}$), associated, correspondingly, to the TPT and MPT potentials, in terms of the deformed operators, can be obtained by expanding the corresponding functions $\arcsin(\hat{y}_{+})$ and $\textrm{arcsinh}(\hat{y}_{-})$ around zero, with $\hat{y}_{+}=\sin(a x_{+})$ and $\hat{y}_{-}=\sinh(a x_{-})$. In doing so, one should be able to construct, for instance, a $x_{\pm}$-representation of the form: 
\begin{equation}
x_{\pm} =  \sqrt{\frac{\hbar}{2\mu \Omega_{\pm}}} \left( \hat{A}^{\dagger}_{\pm}K_{1,\hat{n}}^{(\pm)}+\hat{A}^{\dagger 3}_{\pm} K_{2,\hat{n}}^{(\pm)} + \hat{A}^{\dagger 5}_{\pm}K_{3,\hat{n}}^{(\pm)} + \ldots \right )+hc., \label{eq:displacement}
\end{equation}
where only odd powers of $\hat{A}^{\dagger}(\hat{A})$ are involved owing to the symmetry of the potentials. Explicit formulae for some weighting diagonal operators $K_{i,\hat{n}}^{(\pm)}$'s, written in terms of the $F^{(\pm)}_{\hat{n}}$'s and $G^{(\pm)}_{\hat{n}}$'s, some of them up to second order contributions in $1/\lambda_{\pm}$ in the series expansion of the coordinate and momentum for both TPT and MPT potentials, as well as a brief description of the validity of the approximations involved, are discussed in appendix \ref{appen:a}. For ease of reference, an approximate expression for the diagonal operators $K_{1,\hat{n}}^{(\pm)}$, which will be incorporated into the algebraic procedure presented in the next section, are written down, up to the first first-order $\lambda_{\pm}^{-1}$-prefactor, as follows:
\begin{equation}\fl
K_{1,\hat{n}}^{(\pm)} = F_{\hat{n}}^{(\pm)}\pm \frac{1}{12 \lambda_{\pm}} \left \{ \hat{n}f^{2}_{\pm}(\hat{n})F_{2,\hat{n}}^{(\pm)}+(\hat{n}+1)f^{2}_{\pm}(\hat{n}+1)F_{3,\hat{n}}^{(\pm)}+(\hat{n}+2)f^{2}_{\pm}(\hat{n}+2)F_{4,\hat{n}}^{(\pm)} \right \}, \label{eq:k1n}
\end{equation}
where, in turn, the number-dependent functions $F_{i,\hat{n}}^{(\pm)}$ are explicitly given in the appendix; different contributions of the terms accompanying the prefactor ($1/\lambda_{\pm}$) can be chosen to get a better description of a given observable. For the time being, let us focus primarily on their use in what constitutes the main theme of this work. So then, as per accuracy requirements, knowledge of the different diagonal operators appearing in the above representation would make it possible to undertake any feasible application involving the use of this f-oscillator-based approach, making it a more versatile algebraic tool in considering both types of nonlinearities. Incidentally, in \cite{octa3} a different version of the coordinate and momentum representation was employed in which the diagonal operators were basically stated in terms of the numerical calculation of the respective matrix elements of the variables involved, which was moderately useful to explore some issues concerning the properties of nonlinear coherent states within that context, but it has proven not to be particularly advantageous for specific applications. Accordingly, the present work is also an attempt to surmount such a drawback at least within the weak nonlinear regime. So, P\"oschl-Teller-like  oscillators of the type described by Hamiltonian (\ref{eq:spectrum3}), together with the deformed operators (\ref{eq:fladder}), and the position representation (\ref{eq:displacement}), will be our algebraic tool with a view to modeling qubit-nonlinear-oscillator systems in different coupling regimes. 

\section{Qubit-nonlinear-oscillator model: TPT- and MPT-type nonlinearities} \label{sec:3}

The Hamiltonian model of the composite system we seek to describe, which will encompass either TPT or MPT nonlinearities, is proposed to have the following  structure
\begin{equation}
\hat{H}_{\tiny{\textrm{QNO}}}^{(\pm)}=\hat{H}_{\tiny{\textrm{Q}}}+\hat{H}_{\tiny{\textrm{NO}}}^{(\pm)}+\hat{H}^{(\pm)}_{\tiny{\textrm{I}}},
\end{equation}
where each constituent represents the Hamiltonian of a given two-state system (a qubit), $\hat{H}_{\tiny{\textrm{Q}}}$, the respective nonlinear oscillator, $\hat{H}_{\tiny{\textrm{NO}}}^{(\pm)}$, and the interplay between the subsystems involved, $\hat{H}_{\tiny{\textrm{I}}}$. To be more explicit, 
\begin{eqnarray}
\hat{H}_{\tiny{\textrm{Q}}} & = & -\frac{\hbar \Delta_{0}}{2}\hat{\sigma}_{x}-\frac{\hbar \epsilon}{2}\hat{\sigma}_{z},\\
\hat{H}_{\tiny{\textrm{NO}}}^{(\pm)} & = & \frac{\hat{p}^{2}}{2\mu_{\pm}}+V_{\pm}(\hat{x}), \label{eq:hno} \\
\hat{H}_{\tiny{\textrm{I}}}^{(\pm)} & = & g \hat{\sigma}_{z}\hat{x} .
\end{eqnarray}
Here, $\hat{\sigma}_{x}$ and $\hat{\sigma}_{z}$ are the conventional Pauli operators such that $\hat{\sigma}_{z}|\uparrow \ \rangle=|\uparrow \  \rangle$ ($\hat{\sigma}_{z}|\downarrow \  \rangle=-|\downarrow \  \rangle$), with $\{ |\uparrow \ \rangle, |\downarrow \ \rangle \}$ representing the qubit's two logical (or localized) states. The usual nomenclature of circuit QED has been adopted in which the parameters $\Delta_{0}$ and $\epsilon$ denote, respectively, the energy gap and bias of the qubit; nonetheless, a number of physical systems moving in an effective double well potential through quantum tunneling can be restricted, under certain circumstances, to a two-dimensional Hilbert space (see, for instance, \cite{milena} and references therein) and, therefore, are suitable to be described by said terminology. As regards the oscillator system characterized by the confining potentials $V_{\pm}({\hat{x}})$, $\hat{x}$ and $\hat{p}$ are, respectively, the position and momentum operators; $\mu_{\pm}$ is the respective effective mass; and the conventional coordinate-coordinate coupling between the subsystems involved has also been stated \cite{nori,grifoni}, with $g$ characterizing the strength of it. Making use of the algebraic equivalent of (\ref{eq:hno}) given by (\ref{eq:spectrum3}), together with the coordinate representation of the respective oscillator written in terms of the deformed ladder operators (\ref{eq:displacement}), and retaining only up to the first order terms in the expansion, allows us to put forward the approximate Hamiltonian model of the whole system
\begin{equation} \fl 
\hat{H}_{\tiny{\textrm{QNO}}}^{(\pm)} = -\frac{\Delta_{0}}{2}\hat{\sigma}_{x}-\frac{\epsilon}{2}\hat{\sigma}_{z} + \hbar \Omega_{\pm} \left (\hat{n}\pm \frac{\hat{n}^{2}}{2\lambda_{\pm}}\right)+\hbar \bar{g}\hat{\sigma}_{z} \left(K_{1,\hat{n}}^{(\pm)}\hat{A}_{\pm}+\hat{A}^{\dagger}_{\pm}K_{1,\hat{n}}^{(\pm)}\right),
\end{equation}
where $\bar{g}=g\sqrt{\hbar/2 \mu_{\pm}\Omega_{\pm}}$, and an inessential constant term in the oscillator Hamiltonian has been dropped. We find it convenient to rewrite this Hamiltonian in terms of the eigenstates of $\hat{H}_{Q}$, referred to as the ground and excited states, $|g\rangle=\cos(\theta/2)| \uparrow \  \rangle-\sin(\theta/2)| \downarrow \  \rangle$ and $|e\rangle = \sin(\theta/2) |\uparrow \ \rangle+\cos(\theta/2)|\downarrow \ \rangle$, respectively, whose eigenenergies are $\pm \hbar \Delta_{Q}/2$, with $\Delta_{Q}= \sqrt{\epsilon^{2}+\Delta^{2}_{0}}$ being the energy splitting, and where the parameter $\theta$ weights the relative strength of the $\sigma_{x}$- and $\sigma_{z}$-terms via $\tan \theta=-\Delta_{0}/\epsilon$. In doing so, one gets the equivalent algebraic qubit-nonlinear-oscillator model  
\begin{equation} \fl 
\hat{H}_{\tiny{\textrm{QNO}}}^{(\pm)} = \frac{\hbar \Delta_{Q}}{2} \tilde{\sigma}_{z}+\hbar \Omega_{\pm} \left (\hat{n}\pm \frac{\hat{n}^{2}}{2\lambda_{\pm}}\right)-\hbar \bar{g} \left( \frac{\epsilon}{\Delta_{Q}}\tilde{\sigma}_{z} + \frac{\Delta_{0}}{\Delta_{Q}}\tilde{\sigma}_{x} \right) \left (K_{1,\hat{n}}^{(\pm)}\hat{A}_{\pm}+\hat{A}^{\dagger}_{\pm}K_{1,\hat{n}}^{(\pm)}\right),
\label{eq:Hmodel}
\end{equation}
where $\tilde{\sigma}_{z}=|e\rangle \langle e|-|g\rangle \langle g|$ and $\tilde{\sigma}_{x}=|e\rangle \langle g|+|g\rangle \langle e|$. The reason why only up to the first order terms in the $x_{\pm}$-expansion are kept resides in the fact that the present treatment will be restricted to the lowest-lying region of the energy spectrum and, therefore, to the scenario in which $\langle \Omega_{\hat{n}} \rangle$ is not very far from the qubit's transition frequency $\Delta_{Q}$. Thus, as a first approximation, one-quantum excitations ($\Delta n = \pm 1$) are thought of as being the dominant ones even in the moderate coupling regime; this criterion will be benchmarked against the corresponding numerical solution. Indeed, this algebraic representation encapsulates the nonlinear character of the oscillator itself and its interaction with the qubit via an effective $n$-dependent coupling \`a la P\"oschl-Teller through the diagonal operators $K_{1,\hat{n}}^{(\pm)}$'s being explicitly given by (\ref{eq:k1n}). \\

We now turn our attention to tackle the problem of diagonalizing the above Hamiltonian. In this regard, one can make progress in the weak coupling regime ($\bar{g} \ll \Delta_{Q}$) by applying the so-called rotating-wave approximation (RWA), which is tantamount to keeping the counter-rotating terms out of the interaction part of the Hamiltonian; in the present context, they are identified as those $ \propto |e\rangle \langle g| \hat{A}^{\dagger}$ and $ |g\rangle \langle e| \hat{A}$, thereby enabling us to obtain analytical expressions for quantities of physical interest through using, as usual, the conservation of the total number of excitations. In this coupling regime, where the RWA is generally considered to be valid, some results have already been reported in previous studies concerning algebraically similar systems involving a two-level atom coupled, \`a la Jaynes-Cummings, to an optical cavity field possessing Kerr-type nonlinearities \cite{buzek,sivakumar,martin,octa4}.  However, the effectiveness of such an approximation in the case where the strength of the coupling $\bar{g}$ becomes a significant fraction of $\Delta_{Q}$ ({\it e.g.}, $\bar{g}/\Delta_{Q}>0.2$), which is referred to as the ultrastrong-coupling regime, in addition to the effective role played by the $K_{1,\hat{n}}^{(\pm)}$-diagonal terms in the interaction, seems not to be, at first glance, obvious. In the moderate-coupling regime, for which the ratio $\bar{g}/\Delta_{Q} \sim 0.1 \to  \ \sim 0.2 $, application of the Van Vleck perturbation theory has already been undertaken to cope with the problem of diagonalizing Hamiltonian systems consisting of a qubit coupled to either a harmonic \cite{grifoni1} or anharmonic \cite{grifoni2} oscillator beyond the RWA, the latter being described by the undriven quartic oscillator. The aforesaid perturbation theory shall be adopted in this work, in the moderately strong-coupling regime, by encompassing the two types of nonlinearities that we have been referring to within a single algebraic treatment, and knowing the oscillator's coordinate representation enables us to sidestep, unlike the Duffing oscillator treatment, the use of any additional perturbative approach to the bare nonlinear oscillator. On the other hand, the ultrastrong-coupling regime will be analyzed by entirely numerical means, which is facilitated by the fact of having a fitting algebraic model. \\

On the understanding that the coupling parameter $\bar{g}$ is small enough in the sense that $\bar{g}\ll \Delta_{Q}$, let the interaction part of Hamiltonian (\ref{eq:Hmodel}) be regarded, accordingly, as a sufficiently small perturbation. If so, one can make use of Van Vleck perturbation theory whose first step entails constructing the effective Hamiltonian 
\begin{equation}
\hat{H}_{\tiny{\textrm{eff}}}^{(\pm)}=e^{iS^{(\pm)}}\hat{H}_{\tiny{\textrm{QNO}}}^{(\pm)}e^{-iS^{(\pm)}},
\label{eq:heff}
\end{equation}
where the transformation matrix $S^{(\pm)}$ is such that $\hat{H}_{\tiny{\textrm{eff}}}^{(\pm)}$ possesses the same eigenvalues as $\hat{H}_{\tiny{\textrm{QNO}}}^{(\pm)}$ and the matrix elements of it connect only near-degenerate  states. As a result, such an effective Hamiltonian will be block-diagonal in the basis $\{ |e,n\rangle, |g,n+1\rangle \}$, with each block consisting of a two-by-two matrix associated with manifolds of two quasi-degenerate energy levels of the composite system. On the basis of the prescription outlined in ref. \cite{cohen}, the transformation matrix $S^{(\pm)}$ is calculated up to second order in $\bar{g}$ (the detailed formulas are given in appendix \ref{appen:b}) so as to get the relevant matrix elements of the effective Hamiltonian which is decomposed, accordingly, into form $\hat{H}_{\tiny{\textrm{eff}}}^{(\pm)} = \hat{H}_{\tiny{\textrm{eff}}}^{(\pm,0)}+\hat{H}_{\tiny{\textrm{eff}}}^{(\pm,1)}+\hat{H}_{\tiny{\textrm{eff}}}^{(\pm,2)}$, where, in turn, the second superscript denotes the order of perturbation. Thus,
\begin{eqnarray*}
\langle e,n|\hat{H}_{\tiny{\textrm{eff}}}^{(\pm,1)}|g,n+1\rangle & = & \langle g,n+1|\tilde{H}_{eff}^{(\pm,1)}|e,n\rangle = \hbar \Delta_{n}^{(\pm)}\sqrt{n+1}, \\
\langle e,n|\hat{H}_{\tiny{\textrm{eff}}}^{(\pm,2)}|e,n\rangle & = &  \hbar \left [(n+1)W_{1,n+1}^{(\pm)}-n \left (W_{1,n}^{(\pm)}+W_{0,n}^{(\pm)}\right) \right ],\\
\langle g,n|\hat{H}_{\tiny{\textrm{eff}}}^{(\pm,2)}|g,n\rangle & = & \hbar \left[ -nW_{1,n}^{(\pm)}+(n+1)\left (W_{1,n+1}^{(\pm)}+W_{0,n+1}^{(\pm)}\right ) \right],
\end{eqnarray*}
where 
\begin{eqnarray} 
\Delta_{n}^{(\pm)} & = & - \bar{g}\frac{\Delta_{0}}{\Delta_{Q}}K_{1,n}^{(\pm)}f_{\pm}(n+1), \\ 
W_{1,n}^{(\pm)}  & = &   - K_{1,n-1}^{(\pm)2}f^{2}_{\pm}(n) \frac{ \epsilon^{2}}{\Delta_{Q}^{2}\Omega_{n-1}^{(\pm)}}\bar{g}^{2}, \\ 
W_{0,n}^{(\pm)} & = & - K_{1,n-1}^{(\pm)2}f^{2}_{\pm}(n) \frac{ \Delta_{0}^{2}}{ \Delta_{Q}^{2} \left (\Delta_{Q}+\Omega_{n-1}^{(\pm)}\right )}\bar{g}^{2},
\end{eqnarray}
where, in turn, as stated in the previous section, $\Omega_{n}^{(\pm)}$ is identified as the frequency separation between adjacent energy levels of the respective nonlinear oscillator given by eq. (\ref{eq:frecuencyn}). So, the matrix representation of $\hat{H}_{\tiny{\textrm{eff}}}^{(\pm)}$ in the basis states $|e,n\rangle $ and $|g,n+1\rangle$ can be viewed, for a given value of $n$, as follows: 
\begin{equation} \fl
\hat{H}_{\tiny{\textrm{eff}}}^{(\pm)} \doteq \hbar \left(\begin{array}{cc} \frac{\Delta_{Q}}{2}+\Omega_{\pm} \left(n \pm \frac{n^{2}}{2\lambda_{\pm}} \right) & \sqrt{n+1}\Delta_{n}^{(\pm)}  \\  + (n+1)W_{1,n+1}^{(\pm)}-n \left (W_{1,n}^{(\pm)}+W_{0,n}^{(\pm)}\right) & \\ &  \\ 
\sqrt{n+1}\Delta_{n}^{(\pm)} &  -\frac{\Delta_{Q}}{2}+\Omega_{\pm} \left(n+1\pm \frac{(n+1)^{2}}{2\lambda_{\pm}} \right) \\
& -(n+1)W_{1,n+1}^{(\pm)}+(n+2)\left (W_{1,n+2}^{(\pm)}+W_{0,n+2}^{(\pm)} \right) \end{array} \right).
\end{equation}
The eigenvalues of $\tilde{H}_{eff}^{(\pm)}$ and the corresponding eigenvectors are found to be given, respectively, by  
\begin{eqnarray}
E_{2n+2}^{(\pm)} & = &  \frac{\hbar}{2} \left \{ \eta_{n}^{(\pm)} + \sqrt{\delta_{n}^{(\pm)2}+4(n+1)\Delta_{n}^{(\pm)2}} \right \}, \label{eq:eigenenergy1} \\
E_{2n+1}^{(\pm)} & = &  \frac{\hbar}{2} \left \{ \eta_{n}^{(\pm)} - \sqrt{\delta_{n}^{(\pm)2}+4(n+1)\Delta_{n}^{(\pm)2}} \right \}, \label{eq:eigenenergy2}
\end{eqnarray}
and
\begin{eqnarray}
|\phi^{(\pm)}_{2n+2} \rangle_{\tiny{\textrm{eff}}} & = & \cos \left( \frac{\alpha_{n}^{(\pm)}}{2} \right)|e,n\rangle - \sin \left( \frac{\alpha_{n}^{(\pm)}}{2} \right) |g,n+1\rangle , \\
|\phi^{(\pm)}_{2n+1} \rangle_{\tiny{\textrm{eff}}} & = & \sin \left( \frac{\alpha_{n}^{(\pm)}}{2} \right)|e,n\rangle + \cos \left( \frac{\alpha_{n}^{(\pm)}}{2} \right)|g,n+1\rangle ,
\end{eqnarray}
for $n \ge 0$. Here, the following definitions have been set
\begin{eqnarray*}
\eta_{n}^{(\pm)} & = & \Omega_{n}^{(\pm)}+2\Omega_{\pm} nf^{2}_{\pm}(n+1)+\Delta W_{n,-}^{(\pm)}, \\
\delta_{n}^{(\pm)}  & = & \Delta_{Q} -\Omega_{n}^{(\pm)}+2(n+1)W_{1,n+1}^{(\pm)}-\Delta W_{n,+}^{(\pm)}, \\ 
\tan \alpha_{n}^{(\pm)}  & = &  -\frac{2\sqrt{n+1}\Delta_{n}^{(\pm)}}{\delta_{n}^{(\pm)}},
\end{eqnarray*}
together with
\begin{eqnarray*}
\Delta W_{n,+}^{(\pm)} & = &  (n+2)\left (W_{1,n+2}^{(\pm)}+W_{0,n+2}^{(\pm)}\right )+ n \left (W_{1,n}^{(\pm)}+W_{0,n}^{(\pm)} \right), \\
\Delta W_{n,-}^{(\pm)} & = &  (n+2)\left (W_{1,n+2}^{(\pm)}+W_{0,n+2}^{(\pm)}\right )- n \left (W_{1,n}^{(\pm)}+W_{0,n}^{(\pm)} \right).
\end{eqnarray*}
And, the energy of the ground state $| \phi_{0}^{(\pm)} \rangle_{\textrm{eff}}= |0,g\rangle$ of the effective Hamiltonian is found to be given by the approximate expression
\begin{equation}
E_{0}^{(\pm)} = \hbar \left( -\frac{\Delta_{Q}}{2}+W_{1,1}^{(\pm)}+W_{0,1}^{(\pm)} \right).
\label{eq:cground}
\end{equation}
So, it follows from transformation (\ref{eq:heff}) that the eigenstates of $\hat{H}_{\tiny{\textrm{QNO}}}$ are: $|\phi_{2n+2}^{(\pm)}\rangle = e^{-iS^{(\pm)}} |\phi_{2n+2}^{(\pm)}\rangle_{\tiny{\textrm{eff}}}$, $|\phi_{2n+1}^{(\pm)}\rangle = e^{-iS^{(\pm)}} |\phi_{2n+1}^{(\pm)}\rangle_{\tiny{\textrm{eff}}}$, and $|\phi_{0}^{(\pm)}\rangle =e^{-iS^{(\pm)}} |\phi^{(\pm)}_{0} \rangle_{\tiny{\textrm{eff}}}$; with the last one denoting the respective ground state. 

\begin{figure}[h!]
\begin{center}
\includegraphics[width=7cm, height=4.5cm]{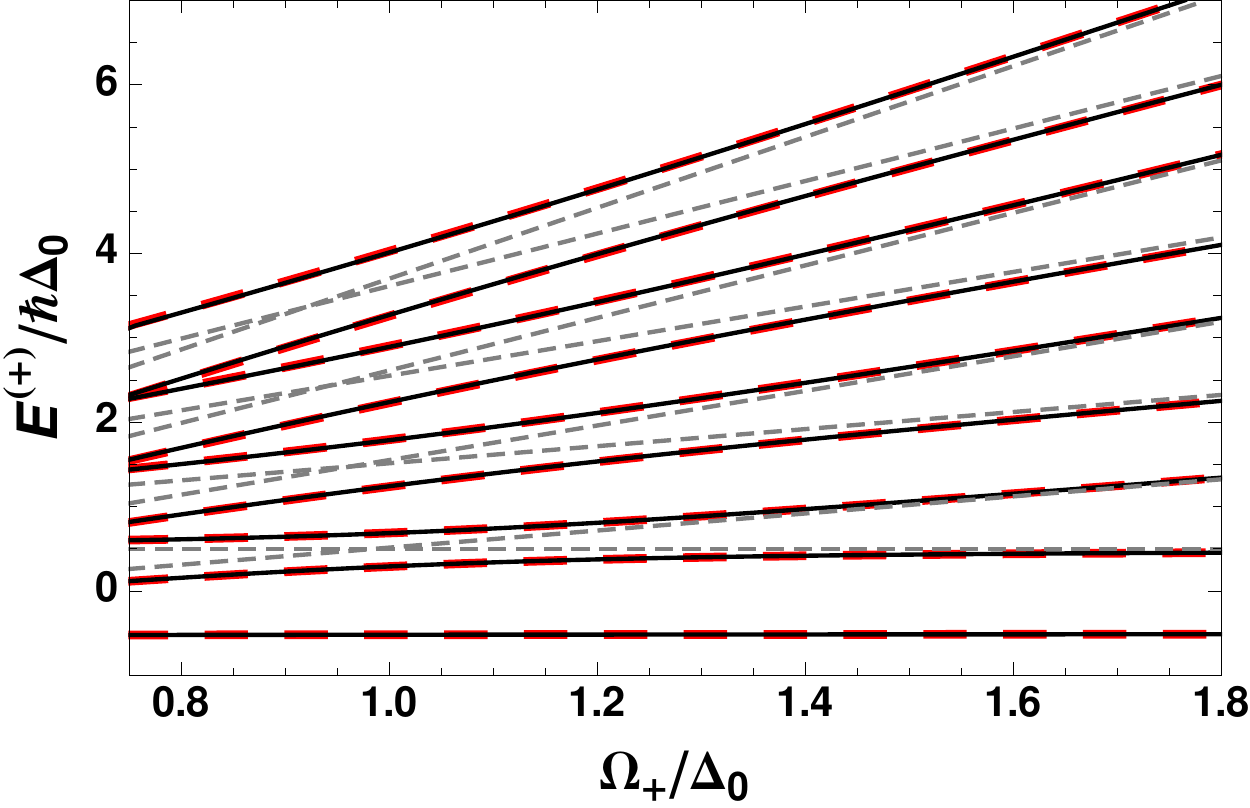} 
\includegraphics[width=7cm, height=4.5cm]{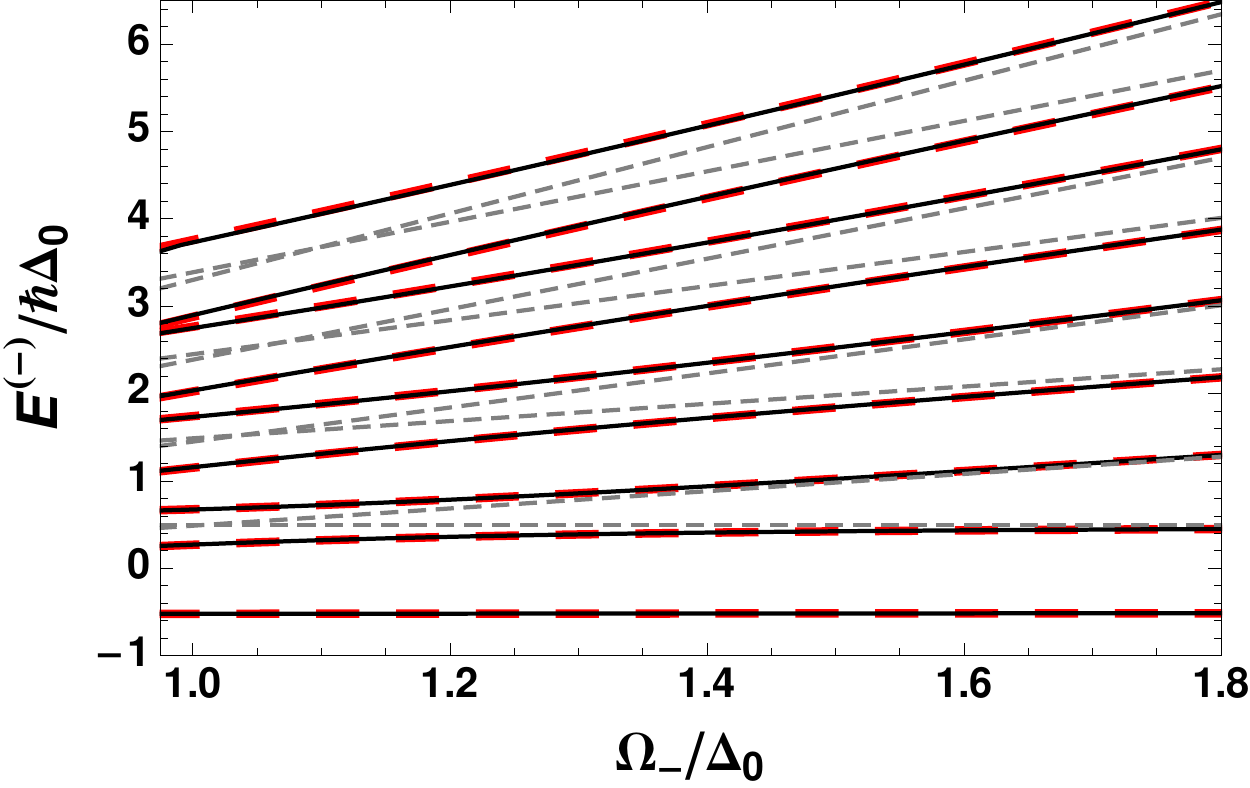} 
\caption{Lowest energy levels of the coupled (red-long-dashed and black-solid lines correspond to the approximate and numerical results, respectively) and uncoupled (grey-dashed lines) qubit-NO system as functions of the reference frequency $\Omega_{\pm}$; left and right panels show, correspondingly, the results concerning the TPT- and MPT-nonlinear systems. The frequencies and energies are plotted in units of $\Delta_{0}$ and $\hbar \Delta_{0}$, respectively. The parameters are: $\epsilon=0$ (unbiased qubit), $\lambda^{-1}_{\pm} = 0.025$ (degree of anharmonicity), and $\bar{g}/\Delta_{0}=0.2$.}
\label{fig:energy-levels}
\end{center}
\end{figure}

This is the set of analytical results within the moderate-coupling regime to second-order Van Vleck perturbation theory in the parameter $\bar{g}$. The corresponding pictorial representation of the eigenvalues (\ref{eq:eigenenergy1}) and (\ref{eq:eigenenergy2}) is shown in Fig. \ref{fig:energy-levels} for the TPT (left panel) and MPT (right panel) nonlinear systems; in the figure, the lowest nine energy levels are plotted as functions of the ratio $\Omega_{\pm}/\Delta_{0}$, for $\epsilon = 0$ and $\lambda_{\pm}^{-1} = 0.025$. In both cases, the outcome of the approximation (red-dashed lines) matches that of the result obtained by numerically diagonalizing (continuous-black lines) a slightly more generalized  version of Hamiltonian (\ref{eq:Hmodel}), which takes into account a $x_{\pm}$-coordinate representation in the interaction part of it including up to third order powers of the deformed ladder operators. That is,
\begin{eqnarray}  
\fl \hat{H}_{\tiny{\textrm{QNO}}}^{(\pm)} & = &  \frac{\hbar \Delta_{Q}}{2} \tilde{\sigma}_{z}+\hbar \Omega_{\pm} \left (\hat{n}\pm \frac{\hat{n}^{2}}{2\lambda_{\pm}}\right) \nonumber \\ 
\fl & & -\hbar \bar{g} \left( \frac{\epsilon}{\Delta_{Q}}\tilde{\sigma}_{z} + \frac{\Delta_{0}}{\Delta_{Q}}\tilde{\sigma}_{x} \right) \left (K_{1,\hat{n}}^{(\pm)}\hat{A}_{\pm}+\hat{A}^{\dagger}_{\pm}K_{1,\hat{n}}^{(\pm)}+K_{2,\hat{n}}^{(\pm)} \hat{A}_{\pm}^{3}+ \hat{A}^{\dagger 3}_{\pm}K_{2,\hat{n}}^{(\pm)} \right),
\label{eq:Hextended}
\end{eqnarray}
where, in turn, the diagonal operators $K_{2,\hat{n}}^{(\pm)}$ are explicitly given by (\ref{eq:diagonalx2}). This fact  confirms that the choice of the one-quantum excitation approach represented by Hamiltonian (\ref{eq:Hmodel}) can be regarded as a quite acceptable first approximation when the strength of the coupling is moderate; entirely numerical calculations presented in the sequel will be based upon the diagonalization of the extended Hamiltonian (\ref{eq:Hextended}). Within the interval shown in the figure, one can discern the doublet structure of the levels taking place near the degeneracy point we have alluding to, which is indicated by the intersections of the energy levels (gray-dashed lines) that  correspond to the uncoupled case ($\bar{g}=0$). One can also see that the position of consecutive crossings is slightly shifted, in frequency, from $\Omega_{\pm}/\Delta_{0}\approx 1$ to the left or right, as the energy increases, depending on the type of nonlinearity; in the weak nonlinear regime, the overall behavior shown in the left panel of the figure is in agreement with the corresponding result obtained in ref. \cite{grifoni2}. And, although the degree of anharmonicity is considered to be small in both nonlinear cases, this gives rise to discernible differences reflected upon the spacing between adjacent energy levels when the strength of the coupling is turned on ($\bar{g}\neq 0$). This fact can approximately be quantified by the following approximate expression
\begin{equation}
E^{(\pm)}_{2n+2}-E^{(\pm)}_{2n+1}  \approx  2\hbar K_{1,n}^{(\pm)}f_{\pm}(n+1)\sqrt{n+1}\bar{g},
\end{equation}
provided that, for the particular case of $\epsilon=0$, the qubit's transition frequency $\Delta_{0}$ be properly tuned to a given $n$-dependent step of the non-equidistant energy ladder of the nonlinear oscillator, in a way such that the relationship $\Delta_{0}-\Omega_{n}^{(\pm)} \approx (n+2)W_{0,n+2}^{(\pm)}+nW_{0,n}^{(\pm)}$ holds; this criterion is only approximately valid for very small values of $\lambda_{\pm}^{-1}$ and those steps belonging to the lowest-lying region of the energy  spectrum. The foregoing relationship, which bears some conceptual (or quantitative) resemblance to the so-called Bloch-Siegert shift \cite{bloch}, is, needless to say, dependent on the degree of anharmonicity via the weighting functions $K_{1,n}^{(\pm)}$ and the deformation $f_{\pm}$ itself.

\section{Ground state of the system and some of its properties} \label{sec:4}

In this section, the energy properties, the uncertainty relations, and the phase space picture associated to the ground state of the composite system, in terms of the parameters $\epsilon$ and $\bar{g}$, but primarily in terms of the latter, are explored and, in parallel, the extent to which the results obtained through  the Van Vleck perturbation theory can be regarded as good approximations is also commented on. Additionally, most of the results pictorially presented in the sequel are also compared with the numerical solution to the linear (or undeformed) case for which $f_{\pm}(\hat{n})=1$. \\

\begin{figure}[h!]
\begin{center}
\includegraphics[width=7cm, height=4.5cm]{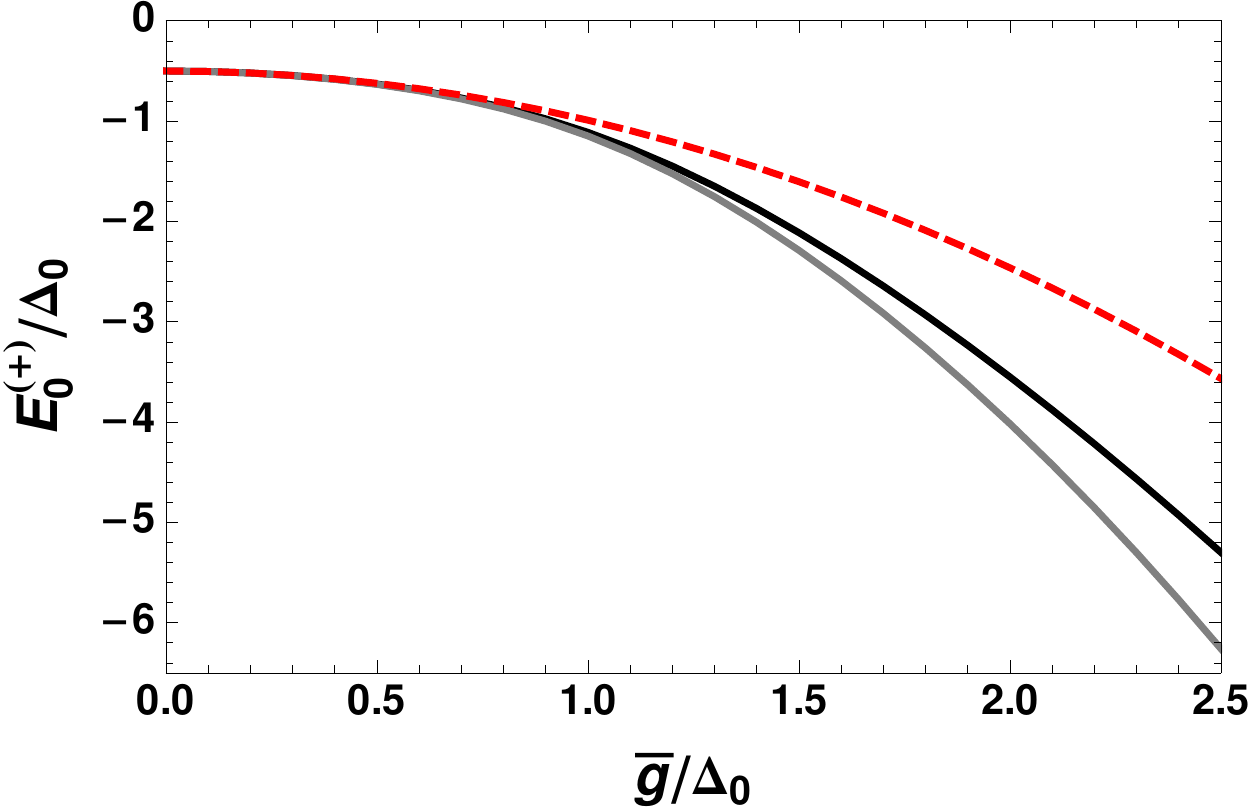} 
\includegraphics[width=7cm, height=4.5cm]{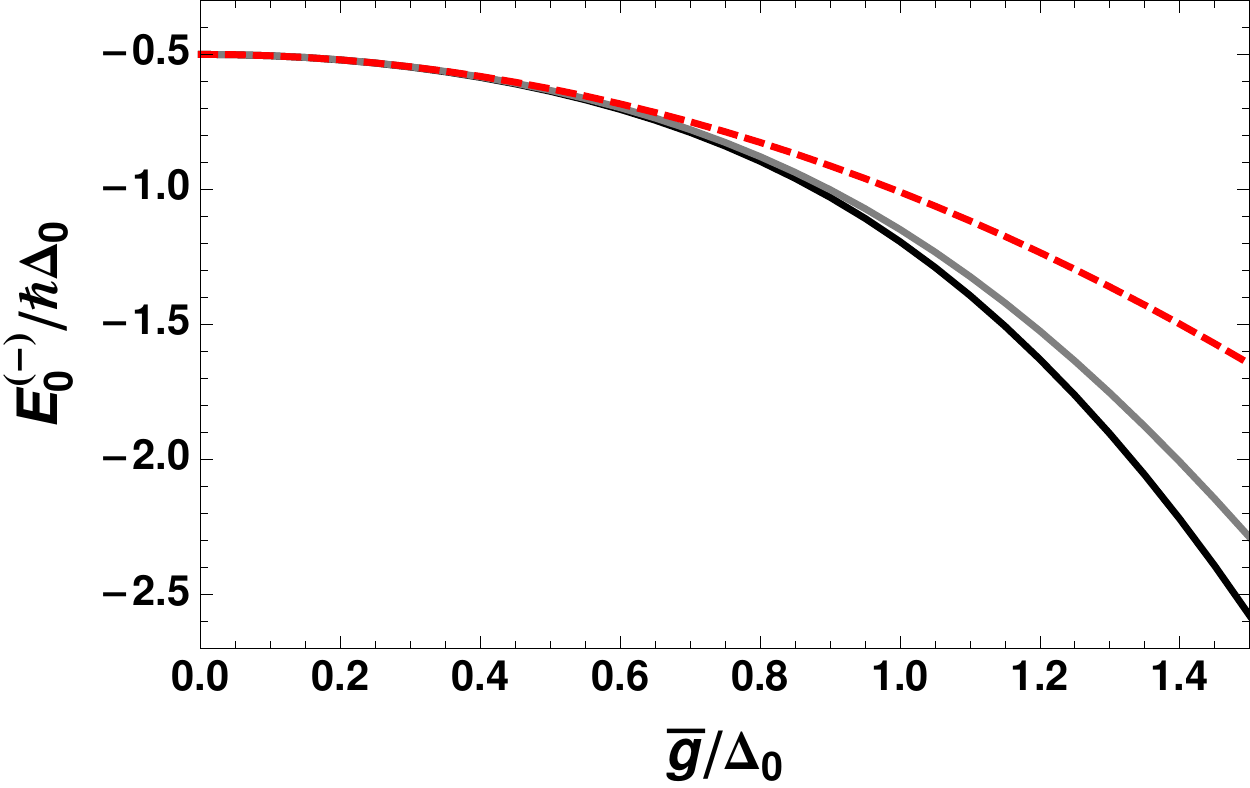} 
\caption{Ground state energy of the composite system as a function of the ratio $\bar{g}/\Delta_{0}$. The figure shows the entirely numerical outcome (black-solid line) coming from a direct diagonalization of the proposed nonlinear model and the approximate result (red-dashed line) given by the Van Vleck perturbation theory, eq. (\ref{eq:ground-state}); for comparison, the numerical result for the harmonic oscillator case $f_{\pm}(n)=1$ (grey-solid line) is also shown. Left and right panels show, respectively, the results corresponding to the TPT- and MPT-nonlinear systems. The parameters are: $\epsilon=0$ (unbiased qubit), $\lambda^{-1}_{\pm} = 0.025$ (degree of anharmonicity) and $\Omega_{\pm}/\Delta_{0}=1$.}
\label{fig:ground-state}
\end{center}
\end{figure}

\begin{figure}[h!]
\begin{center}
\includegraphics[width=7cm, height=4.5cm]{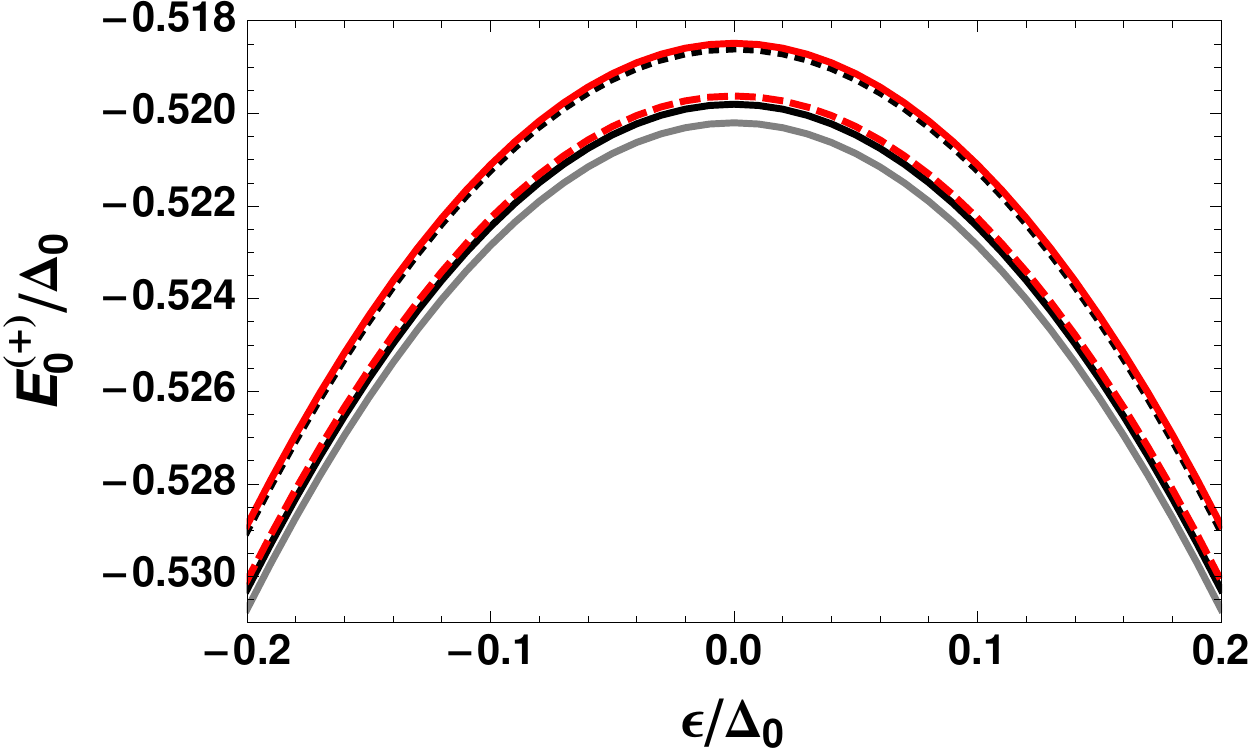} 
\includegraphics[width=7cm, height=4.5cm]{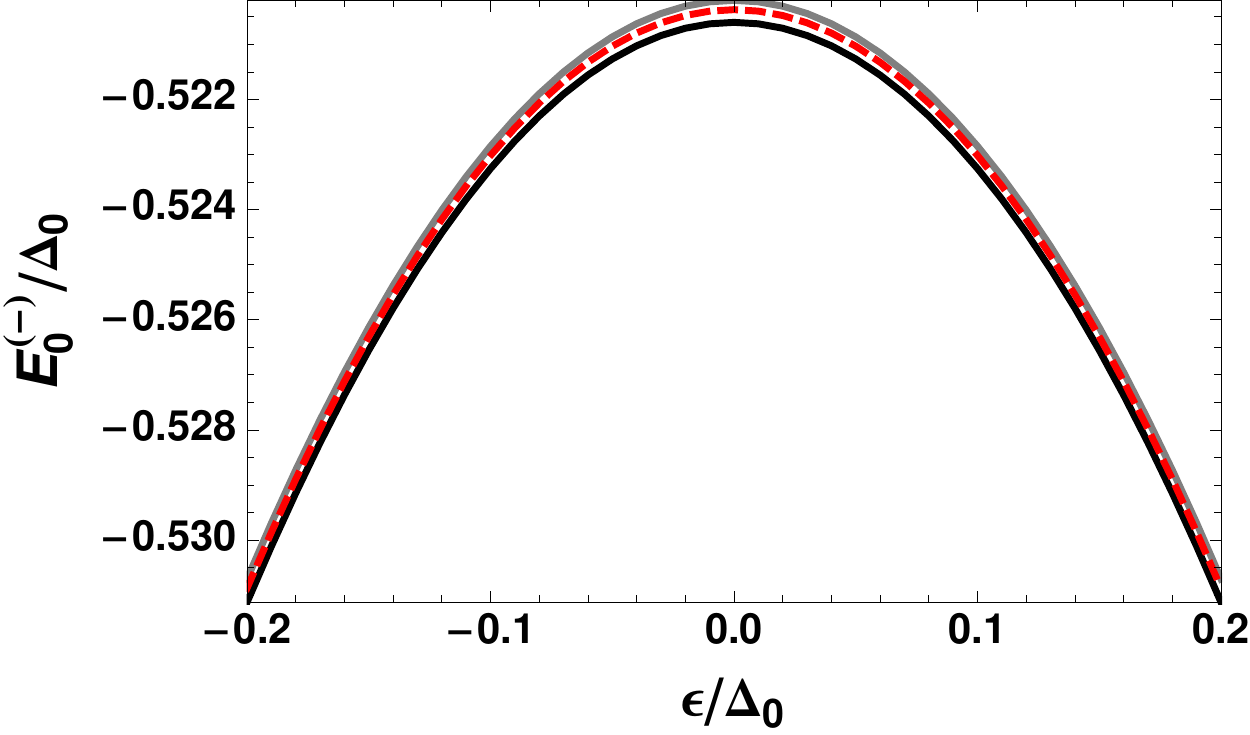} 
\caption{Ground state energy of the composite system as a function of the ratio $\epsilon/\Delta_{0}$. Left panel: Comparison of the entirely numerical results coming from diagonalizing the corresponding nonlinear model with those obtained via the Van Vleck perturbation theory for the TPT-nonlinear case. i) $\lambda_{+}^{-1}= 0.025$ (black-solid line: numerical, red-dashed line: Van Vleck); ii) $\lambda_{+}^{-1}=0.1$ (dotted-black line: numerical, red-continuous line: Van Vleck). Right panel: Corresponding numerical (black-solid line) and approximate (red-dashed line) results for the MPT-nonlinear oscillator with $\lambda^{-1}_{-} = 0.025$. The numerical result for the harmonic oscillator case $f_{\pm}(n)=1$ (grey-solid line) is also shown in both panels as a reference. The remaining parameters are: $\bar{g}/\Delta_{0}=0.2$ and $\Omega_{\pm}/\Delta_{0}=1$.}
\label{fig:gstate-epsilon}
\end{center}
\end{figure}

A first glance at one of above-mentioned properties is provided by briefly examining the ground state energy of the  system, eq. (\ref{eq:cground}), which is rewritten in its more explicit form
\begin{equation}
E_{0}^{(\pm)} = -\hbar \left \{ \frac{\Delta_{Q}}{2}+\left [ \frac{\epsilon^{2}}{\Delta_{Q}^{2}\Omega_{0}^{(\pm)}}+\frac{\Delta_{0}^{2}}{\Delta_{Q}^{2}(\Delta_{Q}+\Omega_{0}^{(\pm)})} \right ] K_{1,0}^{(\pm) 2}f^{2}_{\pm}(1)\bar{g}^{2}  \right \},
\label{eq:ground-state}
\end{equation}
where $K_{1,0}^{(\pm)}$, $f_{\pm}(1)$ and $\Omega_{0}^{(\pm)}$ are, in turn, functions of the anharmonicity parameter $\lambda_{\pm}$. This expression is plotted in figs. \ref{fig:ground-state} and \ref{fig:gstate-epsilon} as a function of the ratios $\bar{g}/\Delta_{0}$ and $\epsilon/\Delta_{0}$, respectively, for TPT (left panel) and MPT (right panel) nonlinear oscillators both. In fig. \ref{fig:ground-state}, one can see that, for $\epsilon =0$, $\lambda^{-1}_{\pm} = 0.025$ and $\Omega_{\pm}/\Delta_{0}=1$, the outcome of both the Van Vleck perturbation theory (red-dashed line) and that of the numerical result corresponding to the harmonic oscillator case  $f_{\pm}(n)=1$ (gray-solid line) seem to offer a quite acceptable description of the ground state energy within the limits of the moderate coupling and the beginnings of the ultra-strong coupling, say, $0<\bar{g}/\Delta_{0} <0.8$, approximately; the effect of a very small anharmonicity is subtle in this interval. For $\bar{g}/\Delta_{0}>1$, the failure of the Van Vleck approximation becomes even more conspicuous than the one of the numerical harmonic case, and discrepancies between the latter the entirely numerical anharmonic result based upon Hamiltonian (\ref{eq:Hextended}) come about, which is slightly more noticeable for the MPT-nonlinear-oscillator case (right panel). Indeed, it is found that the larger coupling strength is, the more evident the nonlinear effects are. Zooming in on the moderate coupling, say $\bar{g}/\Delta_{0}=0.2$, the same equation is plotted in fig. \ref{fig:gstate-epsilon} against the ratio $\epsilon/\Delta_{0}$, together with the corresponding anharmonic and harmonic numerical results. In order to emphasize the role of the anharmonicity in the TPT-nonlinear case (left panel), whose model allows $\lambda_{+}^{-1}$ to acquire slightly larger values than those of its counterpart of the MPT-nonlinear system, the cases $\lambda_{+}^{-1}=0.025$ and $0.1$ are displayed. We see, on the one hand, that the outcome of the nonlinear model departs from the harmonic case insofar as the anharmonicity increases; this tendency corresponds to the black-continuous ($\lambda_{+}^{-1}=0.025$) and dotted ($\lambda_{+}^{-1}=0.1$) lines in the figure for $\lambda_{+}^{-1}=0.025$ and $0.1$, and the approximate results (Van Vleck) are displayed, respectively, in red-dashed and red-solid lines. On the other hand, with regard to the MPT-nonlinear system (right panel), for which $\lambda_{-}^{-1}=0.025$ (black-solid line), the deviation from the linear model turns out to be not so significant in comparison with the one of the previous case and, again, the Van Vleck approach (red-dashed line) gives us a slightly more accurate outcome than the numerical harmonic case ($f_{-}(n)=1$). It is worth emphasizing that, in order for a wider range of values of the ratio $\bar{g}/\Delta_{0}$ to be explored, the numerical results corresponding to all the nonlinear cases are depicted in this  discussion, and subsequently, by considering the extended Hamiltonian (\ref{eq:Hextended}).\\

\begin{figure}[h!]
\begin{center}
\includegraphics[width=7cm, height=4.5cm]{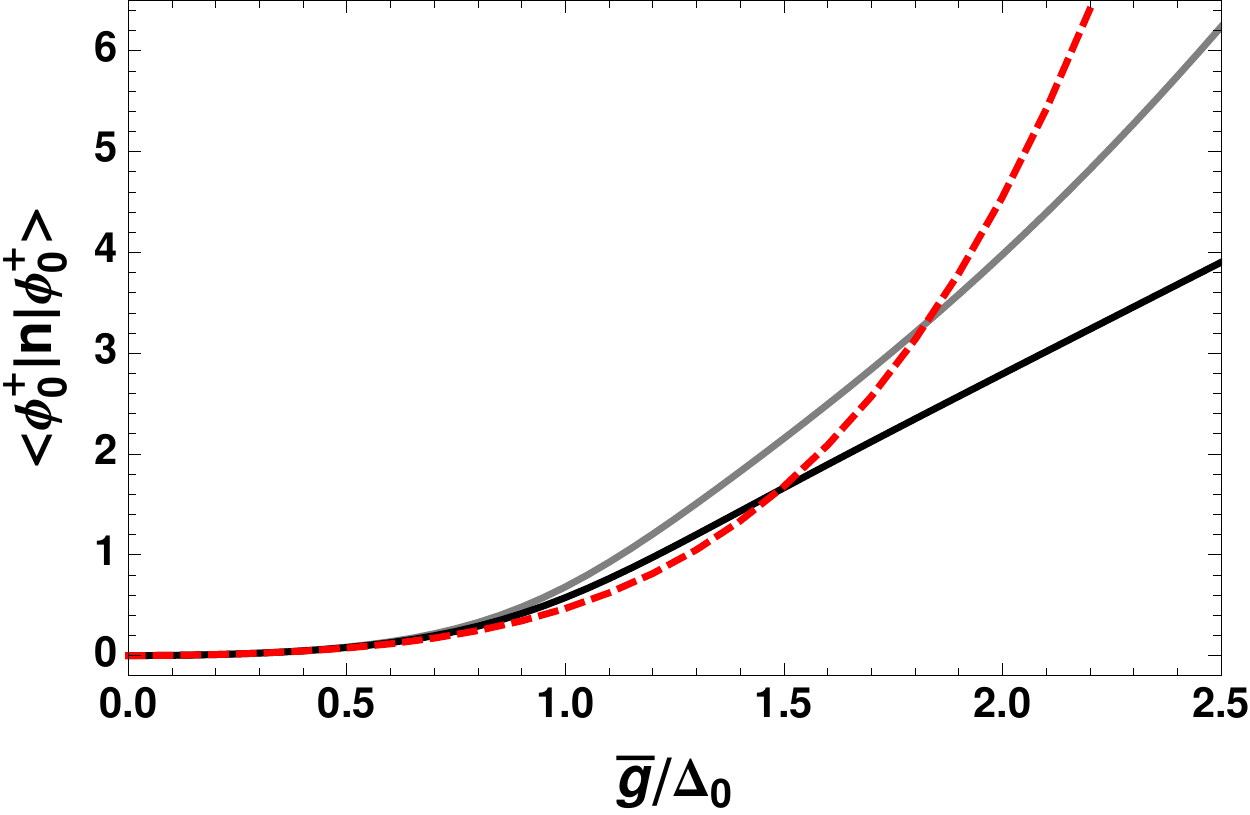} 
\includegraphics[width=7cm, height=4.5cm]{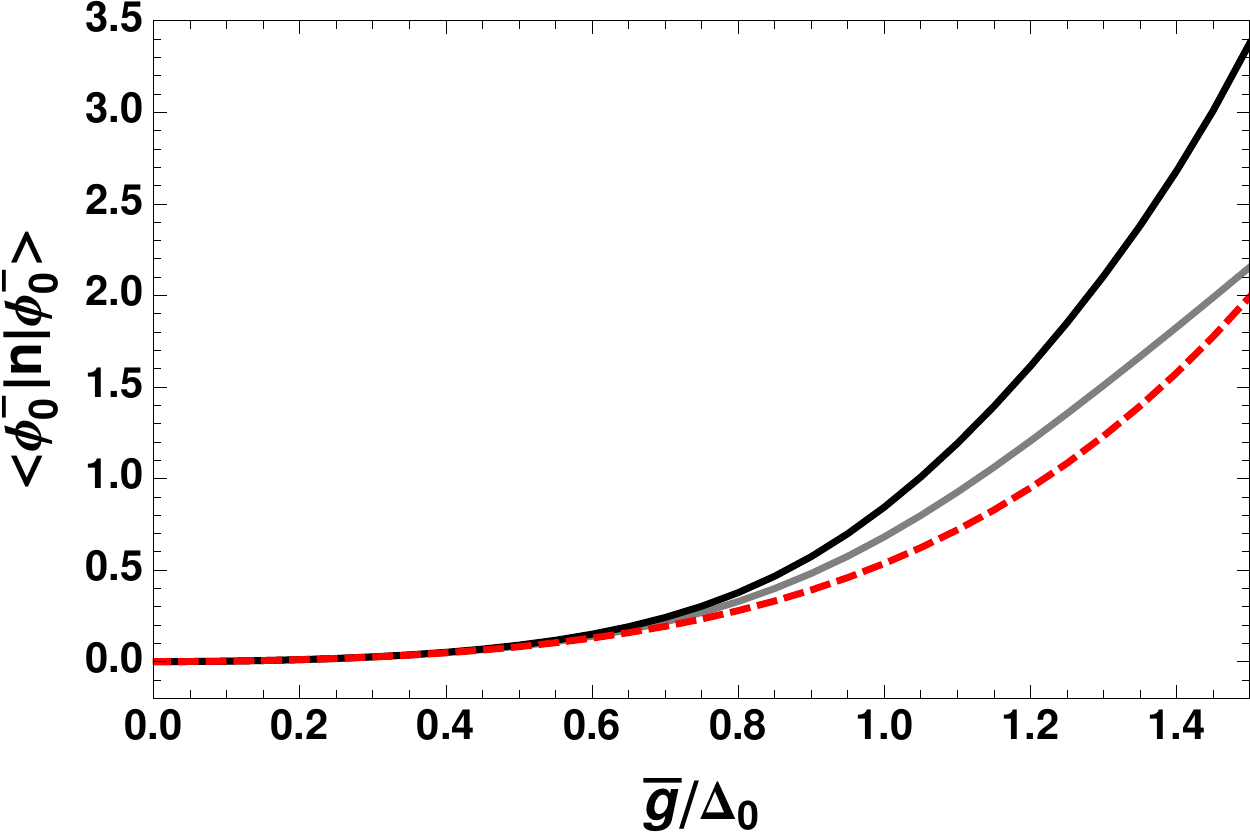} 
\caption{Average number of anharmonic excitations, $\langle \hat{n} \rangle =\textrm{Tr}\{\hat{\rho}_{\tiny{\textrm{NO}}}^{(\pm)} \hat{n} \} $, associated with the system's ground state, as a function of the ratio $\bar{g}/\Delta_{0}$. Left panel: Numerical results (black line), Van Vleck perturbation theory (red-dashed line), harmonic oscillator case (gray line). Right panel:. The parameters are: $\epsilon =0$, $\lambda_{\pm}^{-1}=0.025$ and $\Omega_{\pm}/\Delta_{0}=1$.}
\label{fig:average-n}
\end{center}
\end{figure}

Let us now focus on the statistical properties of the ground state through assessing the degree of occupancy of the oscillator's energy levels by means of the reduced density operator $\hat{\rho}_{\tiny{\textrm{NO}}}^{(\pm)}=\textrm{Tr}_{\tiny{\textrm{Q}}}\{ \hat{\rho}_{\tiny{\textrm{QNO}}}^{(\pm)} \}$, with $\textrm{Tr}_{\tiny{\textrm{Q}}}$ indicating the trace operation over the qubit's degrees of freedom, and $\hat{\rho}_{\tiny{\textrm{QNO}}}^{(\pm)}=| \phi_{0}^{(\pm)}\rangle \langle \phi_{0}^{(\pm)}|$ the ground state density operator of the composite system. If the state $|\phi_{0}^{(\pm)} \rangle $ is the one obtained through the Van Vleck perturbation theory, it follows from it that $|\phi_{0}^{(\pm)} \rangle = \sum_{j}  \{ \langle g,j|e^{-iS^{(\pm)}}|g,0\rangle \  |g,j\rangle +\langle e,j|e^{-iS^{(\pm)}}|g,0\rangle \ |e,j\rangle  \}$, where the weighting coefficients appearing in this expression are evaluated up to second order in $\bar{g}$ according to the general formulae written down in \ref{appen:b}. The Van Vleck approach already discloses the fact that, unlike the well-known weak-coupling limit in which the RWA is meant to be valid, the ground state of the whole system ceases to be the state of zero excitations, $|g,0\rangle$, to become a given coherent superposition of the bare states $\{ |g,j\rangle, |e,j\rangle \}$ possessing a finite number of excitations, thus reflecting the importance of taking into account the counter-rotating terms in the Hamiltonian as the qubit-oscillator coupling increases. From the reduced density operator associated with each nonlinear oscillator, $\hat{\rho}_{\tiny{\textrm{NO}}}^{(\pm)}$, the average number of anharmonic excitations $\langle \phi_{0}^{(\pm)} | \hat{n}| \phi_{0}^{(\pm)} \rangle =\textrm{Tr}\{\hat{\rho}_{\tiny{\textrm{NO}}}^{(\pm)} \hat{n} \} $ can be assessed and the outcome of it is shown in fig. \ref{fig:average-n}, as a function of $\bar{g}/\Delta_{0}$, and compared with the respective numerical results. Again, within the interval $0<\bar{g}/\Delta_{0} < 0.6$, both the numerically computed qubit-harmonic-oscillator model (gray lines) and the Van Vleck approach (red-dashed lines) approximate the outcome of the full numerical solution to the nonlinear model (black lines). By comparison with the harmonic model, it transpires that, in the case of the MPT-nonlinear oscillator (right panel), the larger the strength of the coupling is, the faster the rate at which the anharmonic excitations are created, as opposed to the TPT case (left panel). 

Although the entirely numerical computation of these quantities has been carried out by using up to third order terms in the $x_{\pm}$-representation, it is noted that, for small enough values of the anharmonicity, the one-quantum excitation approach described by Hamiltonian (\ref{eq:Hmodel}) can still provide a good semi-quantitative description with regard to the TPT-based model within the ultrastrong coupling regime. This statement is verified, for instance, by assessing the deviation $\Delta_{\hat{n}}^{(\pm)}=|\langle \phi_{0}^{(\pm)} | \hat{n}| \phi_{0}^{(\pm)} \rangle_{(3)}-\langle \phi_{0}^{(\pm)} | \hat{n}| \phi_{0}^{(\pm)} \rangle_{(1)} |$, where the subscripts denote, respectively, the three- and one-quantum excitation outcomes concerning the average number of quanta. For the sake of brevity, we do not show any plot in this regard. Instead, let us consider the specific set of parameters $\epsilon=0$, $\lambda_{\pm}^{-1}=0.025$ and $\bar{g}/\Delta_{0}=1.5$, where the last one can be regarded as a representative sample within an ultrastrong-coupling segment. In doing so, one obtains that such a deviation reaches the respective values $\Delta_{\hat{n}}^{(+)} \approx 0.03$ and $\Delta_{\hat{n}}^{(-)} \approx 0.3$. As expected, the one-excitation approach starts loosing accuracy as the coupling parameter increases; say, for $\bar{g}/\Delta_{0} =2 $, one gets $\Delta_{\hat{n}}^{(+)} \approx 0.06$ and $\Delta_{\hat{n}}^{(-)} \approx 0.47$. It was verified that the MPT-based model displays the most noticeable deviations when the coupling scale is such that $\bar{g}/\Delta_{0} > 1.25$, whereby the impact of terms proportional to $\hat{A}^{\dagger 3}$ ($\hat{A}^{3}$) in the interaction Hamiltonian must be taken into account.     \\

\begin{figure}[h!]
\begin{center}
\includegraphics[width=7cm, height=4.5cm]{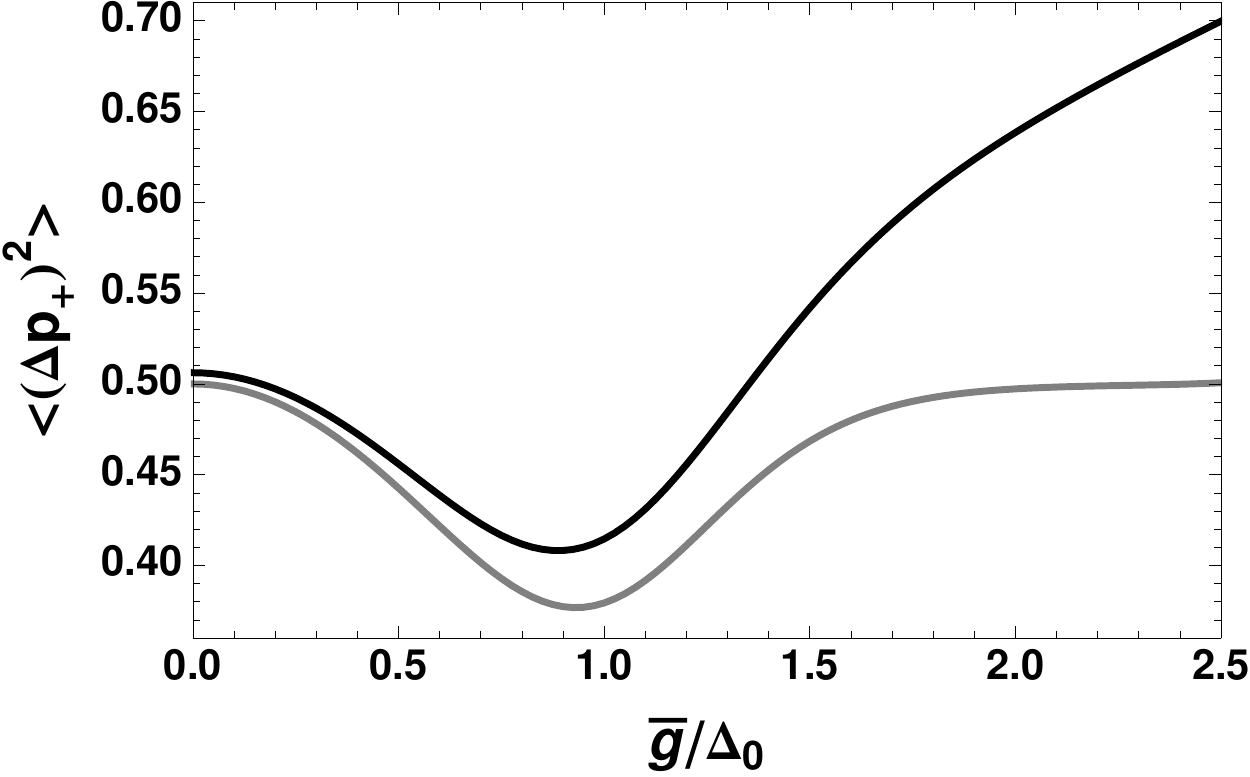} 
\includegraphics[width=7cm, height=4.5cm]{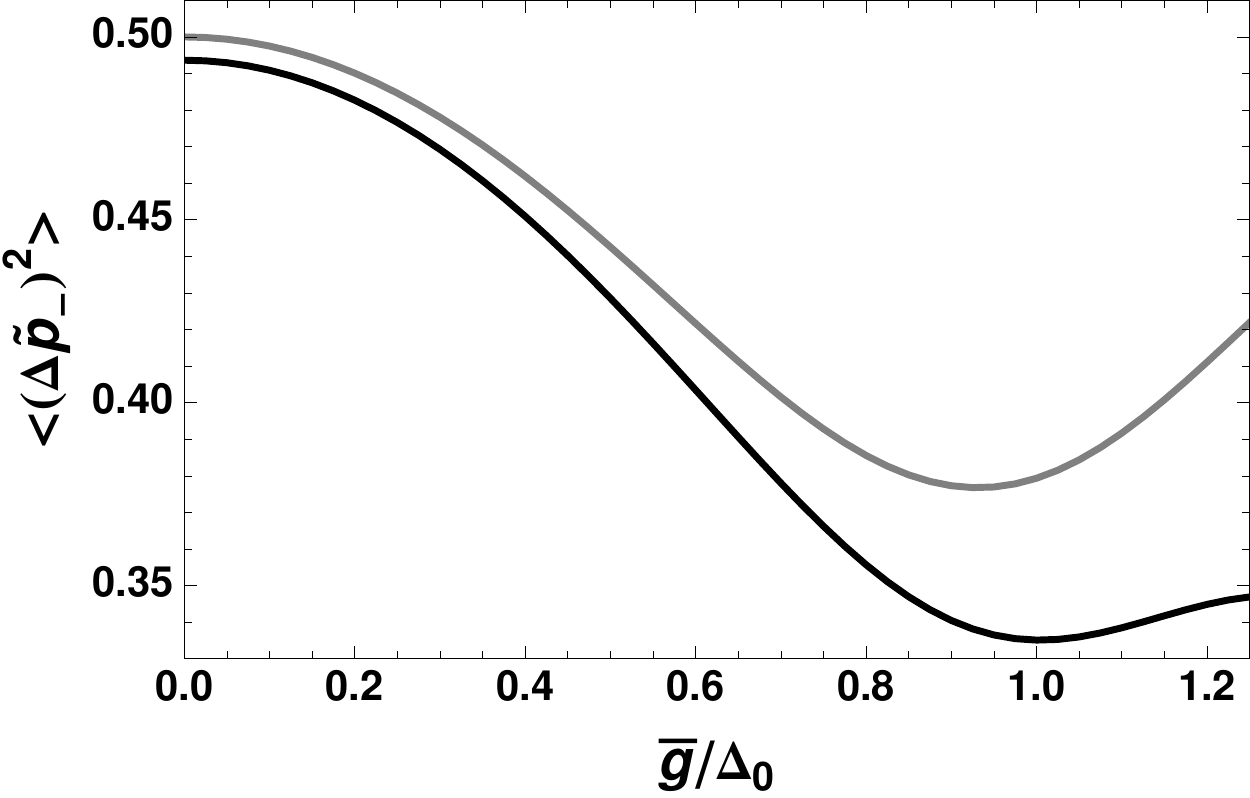} 
\caption{Dispersion in the momentum variable, $\langle (\Delta \tilde{p}_{\pm})^{2} \rangle = \langle \tilde{p}^{2}_{\pm} \rangle-\langle \tilde{p}_{\pm} \rangle ^{2}$, for the ground state as a function of the ratio $\bar{g}/\Delta_{0}$, for the TPT (left panel) and MPT (right panel) nonlinearities. The figure compares the fully numerical outcomes of both the nonlinear (black line) and undeformed, $f_{\pm}(\hat{n})=1$, (gray line) models. The parameters are: $\epsilon =0$, $\lambda_{\pm}^{-1}=0.025$ and $\Omega_{\pm}/\Delta_{0}=1$.}
\label{fig:quadraturep}
\end{center}
\end{figure}

The differences between both nonlinear models become noticeable, particularly in the ultrastrong-coupling regime, in regard to the dispersion in the momentum variable. This feature, assessed by the mean squared deviation $\langle (\Delta \hat{p}_{\pm})^{2} \rangle = \langle \hat{p}_{\pm}^{2} \rangle- \langle \hat{p}_{\pm} \rangle^{2}$, is shown  in fig. \ref{fig:quadraturep} as a function of the ratio $\bar{g}/\Delta_{0}$, where, considering, again, the three-quantum excitation approach, the respective momentum variable $\hat{p}_{\pm}$ is taken to be the approximated one given in  \ref{appen:a} by eq. (\ref{eq:extendp2}); the corresponding numerical results (black lines) are benchmarked against the numeric qubit-harmonic-oscillator case (gray line). For the TPT-like nonlinearity, shown in the left panel of the figure, the ground state of the system can exhibit, for $\bar{g}/\Delta_{0} \in (0.3,1.3)$, the property of squeezing, {\it i.e.}, the variance $\langle (\Delta \hat{p}_{+})^{2} \rangle < 1/2$, but to a slightly lesser degree than the one displayed by its linear counterpart. And, for  longer values of the coupling parameter, the momentum distribution of the ground state displays spreading and departs rapidly from being a minimum uncertainty state. On the other hand, the ground state associated with the MPT case can exhibit squeezing over a much wider $\bar{g}/\Delta_{0}$-interval than the one observed for linear case (gray line), ranging from the weak- to the ultrasong-coupling regime. In this case, due to the limitations of the proposed Hamiltonian model associated to the MPT oscillator, the exploration of this feature was restricted to the interval shown in the figure, in order for the predictions of model to be reliable within the lowest-lying region of the spectrum. \\

\begin{figure}[h!]
\begin{center}
\includegraphics[width=5.5cm, height=5.5cm]{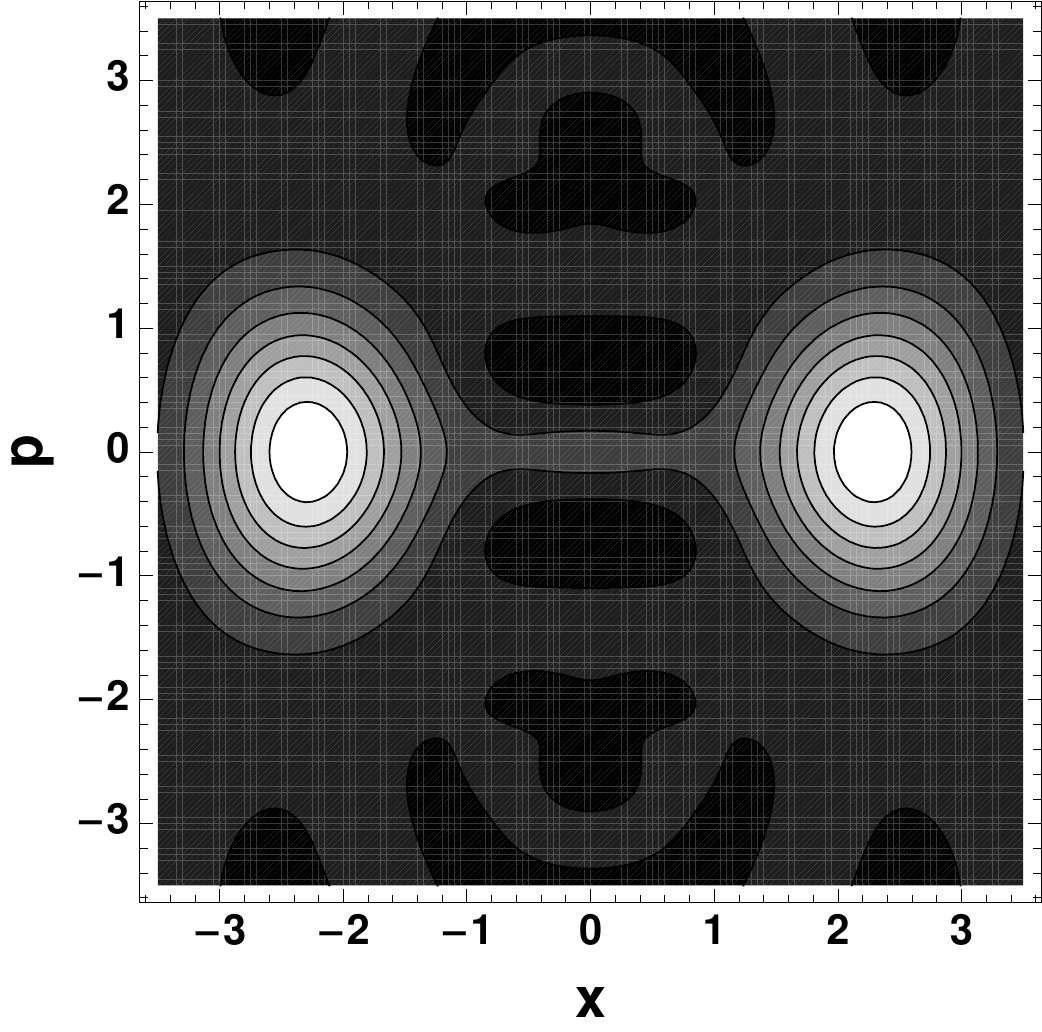} 
\includegraphics[width=5.5cm, height=5.5cm]{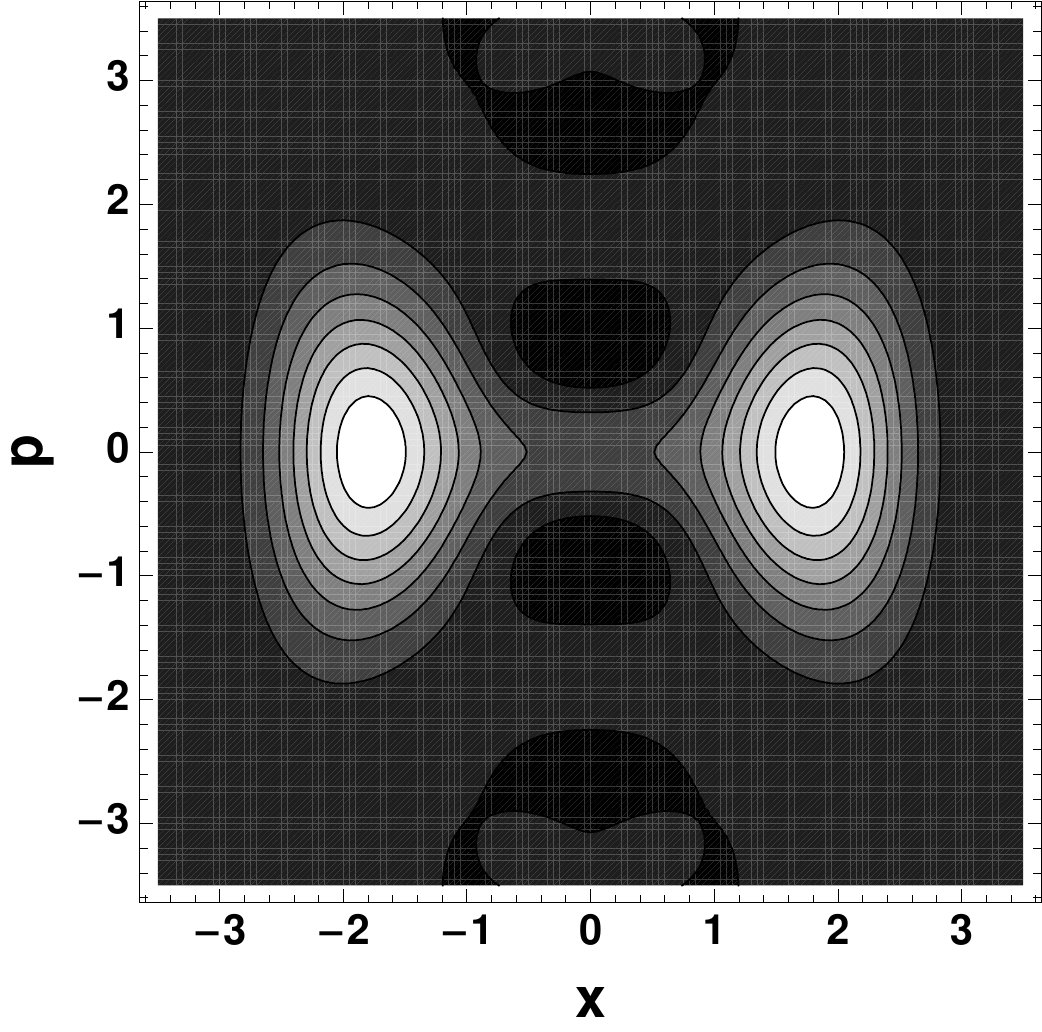} 
\caption{Contour plots of the Wigner function of the oscillator's ground state, in the ultrastrong-coupling regime, $\bar{g}/\Delta_{0}=2$, for the TPT case and two different values of the  anharmonicity parameter: $\lambda_{+}^{-1}=0.025$ (left panel) and $0.08$ (right panel). The dimensionless coordinate $x$ and momentum $p$ are depicted by the abscissa and the ordinate, respectively. The remaining parameters are: $\epsilon=0$ and $\Omega_{+}/\Delta_{0}=1$. Darker and brighter colors indicate, respectively, lower and higher values of the distribution function.}
\label{fig:wigner-functions-tp}
\end{center}
\end{figure}

\begin{figure}[h!]
\begin{center}
\includegraphics[width=5.5cm, height=5.5cm]{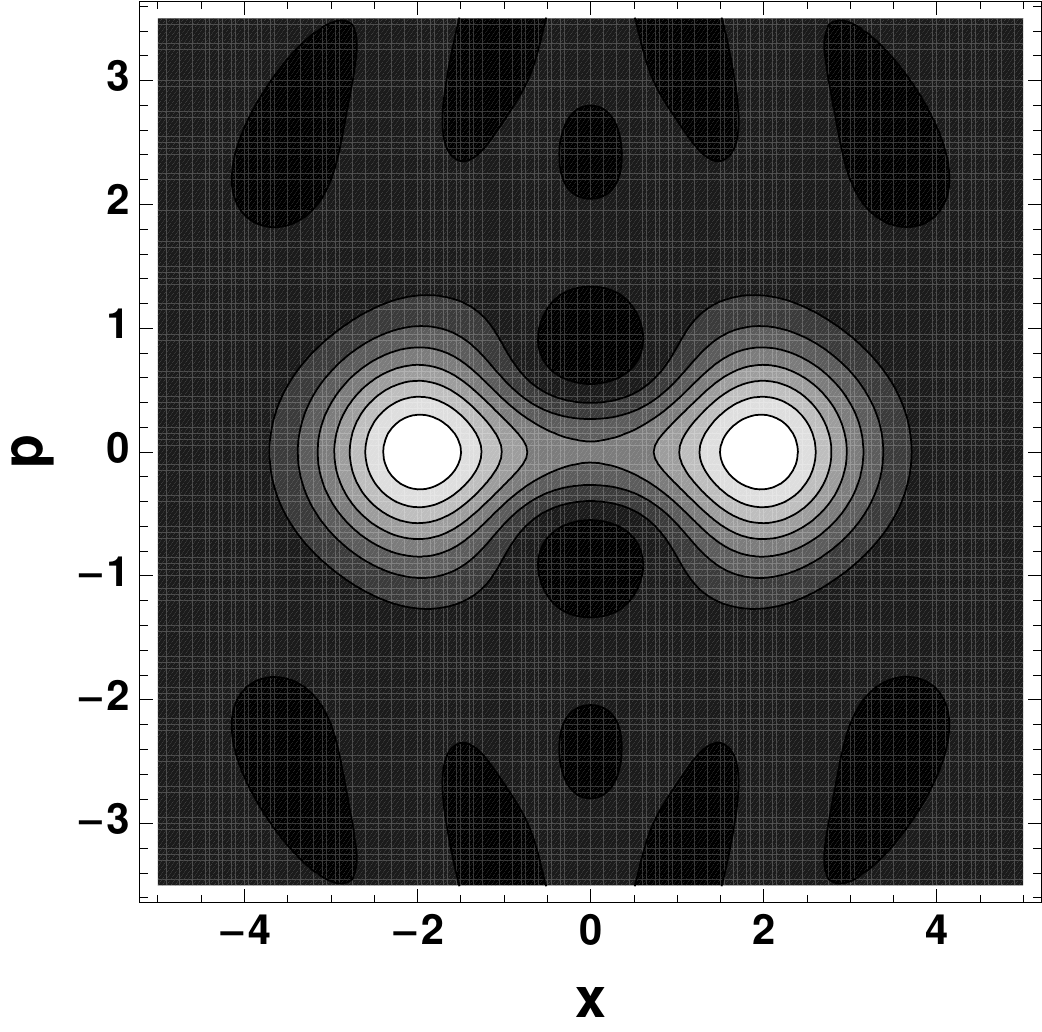} 
\includegraphics[width=5.5cm, height=5.5cm]{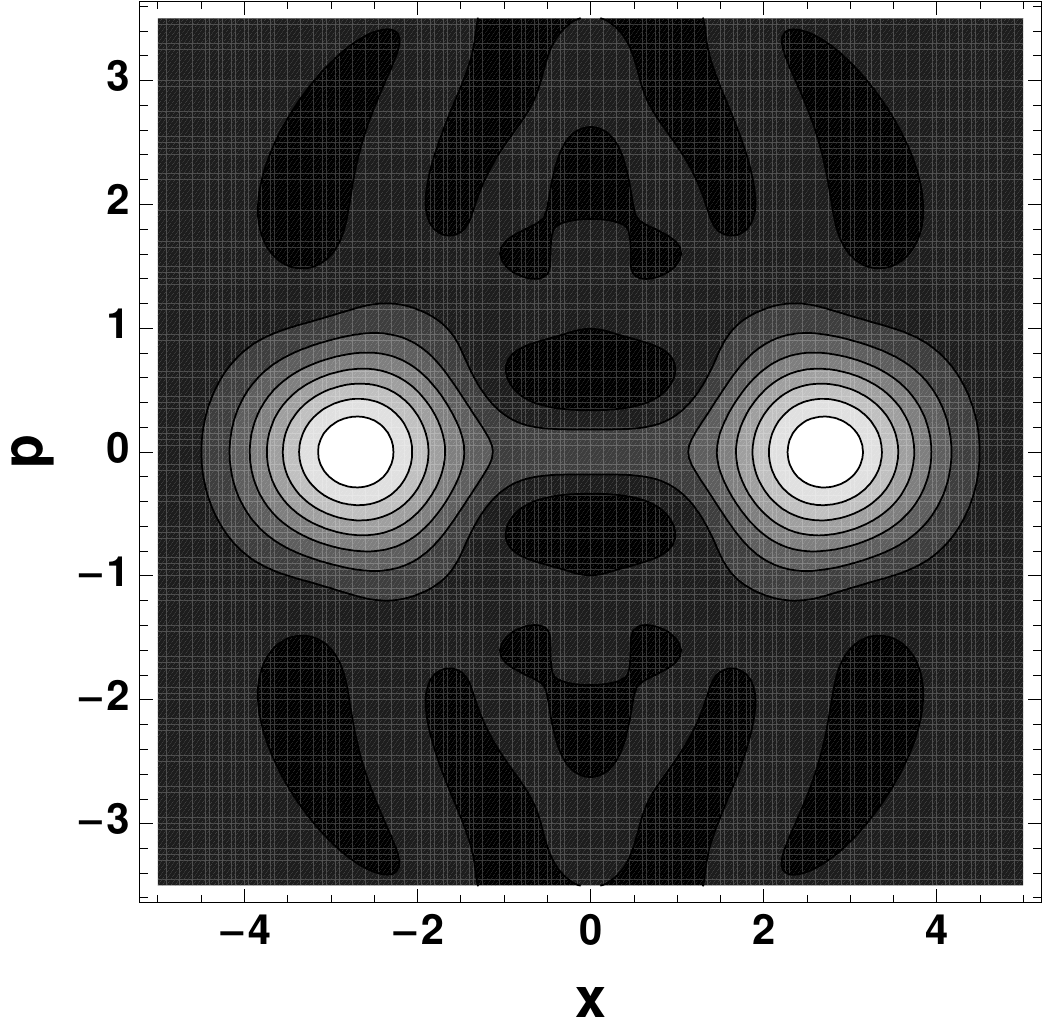} 
\caption{Contour plots of Wigner function of the oscillator's ground state, in the ultrastrong-coupling regime, for the MPT case and two different values of the coupling parameter: $\bar{g}=1.25$ (left panel) and $1.5$ (right panel). The dimensionless coordinate $x$ and momentum $p$ are depicted by the abscissa and the ordinate, respectively. The remaining parameters are: $\epsilon=0$, $\lambda_{-}^{-1}=0.025$ and $\Omega_{-}/\Delta_{0}=1$. Darker and brighter colors indicate, respectively, lower and higher values of the distribution function.}
\label{fig:wigner-functions-mp}
\end{center}
\end{figure}

\begin{figure}[h!]
\begin{center}
\includegraphics[width=7cm, height=4.5cm]{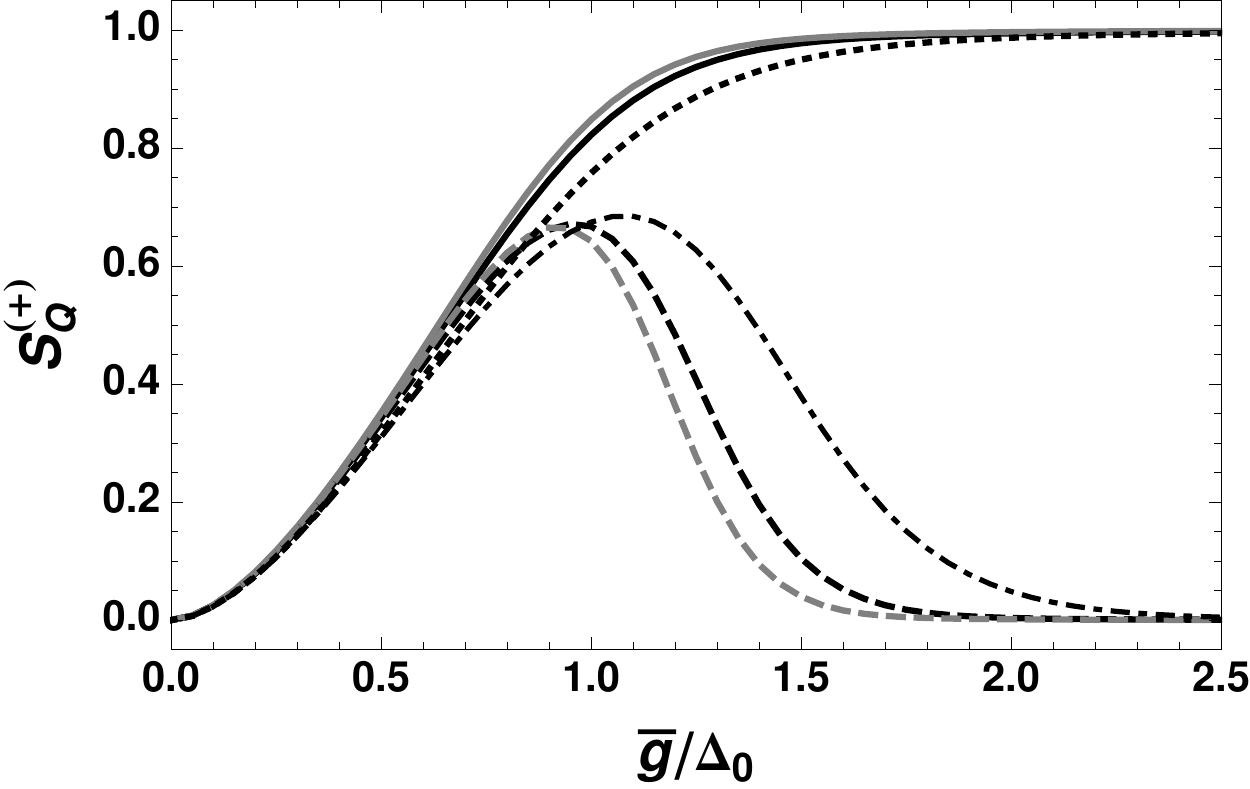} 
\includegraphics[width=7cm, height=4.5cm]{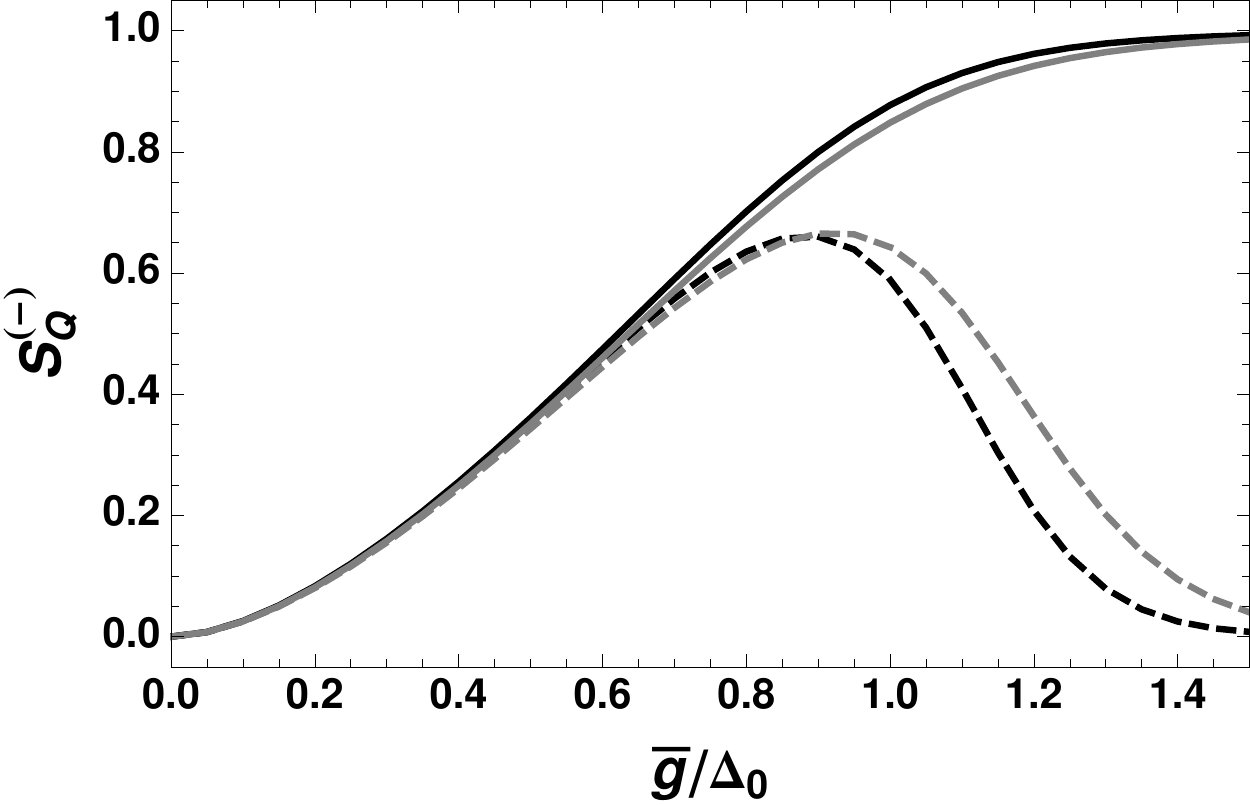} 
\caption{Entropy of the qubit subsystem as a function of the ratio $\bar{g}/\Delta_{0}$. The figures show the full numerical outcome (black lines) for both TPT (left panel) and MPT (right panel) nonlinearities; the linear case, for which $f_{\pm}(n)=1$, continuous- and dashed-gray  line for the unbiased and biased qubit, respectively, is also shown as a benchmark. i) Unbiased qubit ($\epsilon=0$): $\lambda^{-1}_{\pm} = 0.025$ (continuous lines), $\lambda_{+}^{-1}=0.1$ (dotted line, left panel); ii) biased qubit ($\epsilon/\Delta_{0}=0.1$): $\lambda^{-1}_{\pm} = 0.025$ (dashed lines), $\lambda_{+}^{-1}=0.1$ (dotted-dashed line, left panel); $\Omega_{\pm}/\Delta_{0}=1$.}
\label{fig:entropy-qubit}
\end{center}
\end{figure}

The nonclassical character of the ground state can also be revealed by exploring its fingerprint on phase space. To deal with it, the Wigner distribution function is calculated by means of the well-known expression for it \cite{kim}: $W(x,p)=\frac{1}{2 \pi \hbar} \int_{-\infty}^{\infty} \rho (x+y/2,x-y/2)e^{-ipy/\hbar}dy$, where $\rho$ denotes, in our case, the matrix element of the reduced density operator obtained by tracing out the qubit's degrees of freedom, and both $x$ and $p$ are c-numbers. For each nonlinear system, such a matrix element is considered to be given by $\rho(x,x') \to \rho_{\pm}(x,x')= \sum_{m,n}\rho_{m,n}^{(\pm)} \psi_{n}^{(\lambda_{\pm})}(x)\psi_{m}^{(\lambda_{\pm})}(x')$, where the $\rho_{m,n}^{(\pm)}$'s are numerically calculated on the basis of the proposed algebraic model and the $\psi_{j}^{(\lambda_{\pm})}(x)$'s are the wave functions respectively  given by (\ref{eq:wavef1}) and (\ref{eq:wavef2}). How the anharmonicity affects the phase space properties of the state is pictorially represented in figs. \ref{fig:wigner-functions-tp} and \ref{fig:wigner-functions-mp}. By looking at fig. \ref{fig:wigner-functions-tp}, we see the formation of Shr\"odinger-cat-like states for two different values of the anharmonicity parameter, $\lambda_{+}^{-1}=0.025$ (left panel) and $0.08$ (right panel), with $\bar{g}/\Delta=2$ begin fixed in both cases. This effect is somewhat similar the one reported by Ashhab and Nori \cite{nori} in the linear case, save for the fact that the nonlinear feature of our model based on the TPT potential tends to compress, along the coordinate variable, the two coherent constituents of the overall state insofar as the anharmonicity increases while the momentum distribution becomes widespread; the latter property is in accord with the result displayed in the left panel of fig. \ref{fig:quadraturep}. As far as the MPT case is concerned, displayed in fig.  \ref{fig:wigner-functions-mp}, a similar cat-like splitting effect can be seen for $\bar{g}/\Delta_{0}=1.25$ (left panel) and 1.5 (right panel) while the anharmonicity remains fixed at $\lambda_{-}^{-1}=0.025$. Contrasting the present behavior with the previous one emphasizes the tendency to squeeze the momentum distribution of the oscillator (which is also compatible with the result displayed in the right panel of fig. \ref{fig:quadraturep}, besides giving rise to what seems to be a richer, and barely discernible, interference pattern in the vicinity of both coherent hills. It is noted in passing that this peculiar coherent behavior of the ground state disappears for the case of a biased qubit, {\it i.e.,} for $\epsilon \neq 0$, regardless of the smallness of such a parameter. In this regard, it transpires that the appearance of such cat-like fingerprint on phase space turns out to be closely related to the degree of mixture (or entanglement) between the oscillator and the qubit. In fact, it is worth commenting that in the case of a linear oscillator interacting with a qubit within the so-called deep strong coupling regime, theoretical predictions about the emergence of Schr\"odinger-cat-like entangled states have already been confirmed, experimentally, in the context of superconducting circuits \cite{yoshihara}; and recent theoretical work in this vein has also been presented in \cite{zhang1}. So, invoking the theorem of Araki and Lieb \cite{araki} permits us to quantify the degree of entanglement between the nonlinear oscillator and the qubit by means of the entropy of the latter, using the standard expression $S_{\tiny{\textrm{Q}}}=-\textrm{Tr} \{ \hat{\rho}_{\tiny{\textrm{Q}}} \log_{2} \hat{\rho}_{\tiny{\textrm{Q}}} \}$, where $\hat{\rho}_{\tiny{\textrm{Q}}}$ is the reduced density operator obtained by tracing out the oscillator's degrees of freedom, which is  illustrated in fig. \ref{fig:entropy-qubit}, as a function of $\bar{g}/\Delta_{0}$, and whose value ranges from 0, for a pure state, to 1, for a maximally mixed state. We see, in the left panel, that, for $\bar{g}/\Delta_{0}=2$ and $\epsilon=0$, the qubit-oscillator entanglement has already reached its maximum in the TPT case, just where the aforementioned phase-space conduct takes place, and almost the same can be said for MPT model (right panel) within the approximate interval $1.2 \le \bar{g}/\Delta_{0} \le 1.5$. For a biased qubit, on the other hand, say $\epsilon/\Delta_{0}=0.1$, the system is able to reach some degree of mixture that subsequently starts up being undermined by the increment of strength of the qubit-oscillator coupling. Deviations from the linear model (gray line) are also highlighted in the figure. 

\section{Concluding remarks} \label{sec:5}

The use of a fitting P\"oschl-Teller-potential-based scheme, in conjunction with the f-deformed oscillator description, was proposed to undertake the algebraic and/or numerical study of a nonlinear version of the ubiquitous qubit-oscillator model, in different coupling regimes, without considering the rotating-wave approximation. It is found that such an algebraic structure, alternative to the quartic (Duffing) oscillator treatment, is able to enables us to capture, within a single algebraic framework, the essence of hard and soft nonlinearities modeled, respectively, on the basis the corresponding TPT- and MPT-oscillator algebras. In the scenario where the strength of the coupling between the qubit and the nonlinear oscillator is considered to be approximately within the moderate regime, say,  $\bar{g}/\Delta_{\tiny{\textrm{Q}}} \in (0.1,0.2)$, such an approach has proven to be useful in deriving approximate analytical expressions for the eigenenergies and eigenstates of the whole Hamiltonian embracing the aforementioned two nonlinearities, into a single scheme, with the help of Van Vleck perturbation theory. The proposed model is also considered to facilitate the numerical analysis of the composite system in a landscape segment of the ultrastrong-coupling regime ({\it e.g.}, $ \bar{g}/\Delta_{\tiny{\textrm{Q}}} \sim 0.5 \to 2.5$ and $0.5 \to 1.5$ for the TPT and MPT cases, respectively) in which the perturbative approach breaks down. One of the main results of this exploration was that the ground state associated with both qubit-nonlinear-oscillator systems turns out to be highly nonclassical in terms of its squeezing properties in the momentum variable, with the MPT system displaying a higher degree of squeezing in the earlier stages of the ultrastrong-coupling regime than that of the linear model, as well as exhibiting a high degree of mixture (or entanglement) between the subsystems involved. Additionally, unlike the standard linear case where no spreading occurs on phase space, the present model foster the formation of deformed Shr\"odinger-cat-like coherent structures. So, having introduce the approximate model, and its limitations, one can proceed to undertake the exploration of the dynamics, both in the non-dissipative and dissipative cases, of the composite system. Particularly, on the basis of the present treatment, the robustness of coherent effects that could emerge during the evolution of this class of systems against of energy losses and/or dephasing are due to be treated elsewhere.  \\

As far as the deformed oscillator model itself is concerned, it is worth highlighting that the present algebraic treatment is a proposal that strives to go beyond representing the dynamical variables, the coordinate and momentum operators, as mere linear combinations of the deformed ladder operators, say, $x\sim \hat{A}+\hat{A}^{\dagger}$ (as is usually done in the literature), which is no longer a faithfully accurate description, as pointed out in \ref{appen:a}, for the TPT-like oscillator case, even if we restrict our treatment to the lowest-lying region of the spectrum and small nonlinearities. So, a more accurate approach has to embody the proper dependence of possible energy transitions on the level of excitation of a system having a non-equidistant energy structure. This was accomplished through fitting weighting diagonal operators that reflect, at least approximately, the genuine nonlinear character of the system being analyzed. The proposed oscillator models themselves, albeit modest, are also expected to be sufficiently versatile to be exploited as a reliable algebraic tool in contexts where the role of certain anharmonicity corrections  could be essential for the description of more realistic systems, situations in which the prevailing linear harmonic oscillator approach may not be entirely satisfactory (a small step in this direction was made in \cite{jauregui} apropos of the open dynamics of a Morse-like oscillator interacting with a thermal field in the Markovian regime). Along this line of research, one can quote the work of Tanimura {\it et al.} \cite{okumura,tatsuchi} in which the analysis of the evolution of inter-molecular vibrations in liquid environments, described by the so-called non-Markovian Brownian oscillator model, calls for a fitting algebraic scheme for the quantum mechanical description of nonlinear system-bath interactions. In this regard, properly adapted deformed algebras open up the possibility of putting forward their use in modeling the aforesaid dynamic processes. 

\ack I want to thank professor J. R\'ecamier for his hospitality at ICF, UNAM. Thanks also to DGAPA-UNAM for partial support through project IN113016.


\appendix

\section{Expansion of $\hat{x}$ and $\hat{p}$ in terms of deformed operators}\label{appen:a}

In this appendix, an approximate representation of the pair of position $\hat{x}$ and momentum $\hat{p}$ variables associated with P\"oschl-Teller oscillators is proposed within the scheme of the f-deformed algebra, a procedure that was inspired by the work of Lemus and Bernal \cite{lemus}. Originally approached from the viewpoint of the su(2) one-dimensional vibron model, the authors of that work put forward their algebraic method for obtaining an approximate realization of the aforementioned physical variables for the MPT potential solely. Here, the same prescription is applied to the TPT potential and the results thus obtained are found to be much more accurate than the ones of its MPT counterpart; this is one of the results of the present work. \\

So, by virtue of the algebraic equivalence between the standard factorization method and the f-deformed oscillator approach within which the deformation function (\ref{eq:fdeformation}) is established, it makes sense to think of the model Hamiltonian (\ref{eq:spectrum3}) and the set of operators $\{ \hat{A}_{\pm}, \hat{A}^{\dagger}_{\pm},\hat{n}\}$ together as a fitting algebraic scheme for describing oscillatory systems that partake of the P\"oschl-Teller-like nonlinear features. As highlighted in section \ref{sec:2}, this fact justifies the symbolic  correspondence $\hat{b}_{\pm}$ ($\hat{b}^{\dagger}_{\pm}$) $\to$ $\hat{A}_{\pm}$ ($\hat{A}^{\dagger}_{\pm}$) that enables us to derive an approximate representation of the coordinate $x$ and/or the momentum $\hat{p}$ in terms of the deformed operators. To this end, we use the technique of inverting the differential form of the corresponding ladder operators \cite{lemus} applied to both TPT and MPT potentials. In doing so, we extract the coordinate $x$ and the derivative $\frac{d}{dx}$ from (\ref{eq:diferential1}) and (\ref{eq:diferential2}) to get, on the one hand, for the TPT potential, the following relationships  
\begin{eqnarray}
\sin(ax) & = & \frac{1}{\sqrt{2\lambda_{+}}} \left (\hat{A}^{\dagger}_{+}F_{\hat{n}}^{(+)}+\hat{A}_{+} G_{\hat{n}}^{(+)} \right), \label{eq:rel1}\\
\frac{d}{dx} & = & \sqrt{\frac{\lambda_{+}}{2}}\frac{a}{\cos(ax_{+})} \left( \hat{A}_{+}H_{\hat{n}}^{(+)}-\hat{A}^{\dagger}_{+} Q_{\hat{n}}^{(+)}\right), \label{eq:rel2}
\end{eqnarray}
and, on the other hand,  
\begin{eqnarray}
\sinh(ax)  & = &  \frac{1}{\sqrt{2\lambda_{-}}} \left (\hat{A}^{\dagger}_{-}F_{\hat{n}}^{(-)}+\hat{A}_{-} G_{\hat{n}}^{(-)} \right), \\
\frac{d}{dx} & = & \sqrt{\frac{\lambda_{-}}{2}}\frac{a}{\cosh(ax_{-})} \left( \hat{A}_{-}H_{\hat{n}}^{(-)}-\hat{A}^{\dagger}_{-} Q_{\hat{n}}^{(-)}\right), \label{eq:rel3}
\end{eqnarray}
for the MPT potential, where we have set the corresponding number-operator dependent functions
\begin{eqnarray*}
F_{\hat{n}}^{(\pm)} & = & \frac{1}{\sqrt{(1\pm\frac{\hat{n}}{\lambda_{\pm}})(1\pm\frac{\hat{n}+1}{\lambda_{\pm}})}}, \label{eq:funcf} \qquad G_{\hat{n}}^{(\pm)}  =  \frac{1}{\sqrt{(1\pm \frac{\hat{n}}{\lambda_{\pm}})(1\pm\frac{\hat{n}-1}{\lambda_{\pm}})}}, \\
H_{\hat{n}}^{(\pm)} & = & \sqrt{\frac{\lambda_{\pm}\pm \hat{n}}{\lambda_{\pm}\pm \hat{n}\mp 1}}, \qquad Q_{\hat{n}}^{(\pm)}  =  \sqrt{\frac{\lambda_{\pm}\pm \hat{n}}{\lambda_{\pm}\pm \hat{n}\pm 1}}.
\end{eqnarray*}
On setting the shorthand variables
\begin{eqnarray}
\hat{y}_{+} \equiv \frac{1}{\sqrt{2\lambda_{+}}} \left (\hat{A}^{\dagger}_{+}F_{\hat{n}}^{(+)}+\hat{A}_{+} G_{\hat{n}}^{(+)} \right) = \sin(ax), \label{eq:def} \\
\hat{y}_{-} \equiv \frac{1}{\sqrt{2\lambda_{-}}} \left (\hat{A}^{\dagger}_{-}F_{\hat{n}}^{(-)}+\hat{A}_{-} G_{\hat{n}}^{(-)} \right) = \sinh(ax), \label{eq:def2}
\end{eqnarray}
we proceed to expand the following functions about zero 
\begin{eqnarray}
\arcsin (\hat{y}_{+}) & = & \hat{y}_{+}+\frac{\hat{y}^{3}_{+}}{6}+\frac{3\hat{y}^{5}_{+}}{40} +\ldots , \label{eq:approx1} \\ 
\textrm{arcsinh}(\hat{y}_{-}) & = & \hat{y}_{-}-\frac{\hat{y}^{3}_{-}}{6}+\frac{3\hat{y}^{5}_{-}}{40} +\ldots, \label{eq:approx2}\\ 
\frac{1}{\sqrt{1\mp \hat{y}^{2}_{\pm}}} & = & 1\pm \frac{\hat{y}^{2}_{\pm}}{2} + \frac{3\hat{y}^{4}_{\pm}}{8} +\ldots . \label{eq:exp2}
\end{eqnarray}
So, by substituting (\ref{eq:def}) and (\ref{eq:def2}) into (\ref{eq:approx1})-(\ref{eq:exp2}), and, in turn, (\ref{eq:exp2}) into (\ref{eq:rel2}) and (\ref{eq:rel3}), and retaining only up to the first three terms in the above expansions, we arrive, after a long algebra, at approximate expressions for the variables $\hat{x}_{\pm}$ and $\hat{p}_{\pm}$ of the form
{\setlength\arraycolsep{2pt}
\begin{eqnarray}
\fl \hat{x}_{\pm} & \approx & \sqrt{\frac{\hbar}{2\mu \Omega_{\pm}}} \left \{ \hat{A}^{\dagger}_{\pm}F_{\hat{n}}^{(\pm)}\pm \frac{1}{12\lambda_{\pm}}\left (\hat{A}^{\dagger 3}_{\pm}F_{1,\hat{n}}^{(\pm)}+\hat{A}^{\dagger 2}_{\pm}\hat{A}_{\pm}F_{2,\hat{n}}^{(\pm)}+\hat{A}^{\dagger}_{\pm}\hat{A}_{\pm}\hat{A}^{\dagger}_{\pm}F_{3,\hat{n}}^{(\pm)}+\hat{A}_{\pm}\hat{A}^{\dagger 2}_{\pm}F_{4,\hat{n}}^{(\pm)} \right)+ \ldots \right \}  +  hc., \label{eq:extendx} \\
\fl \hat{p}_{\pm} & \approx & i\sqrt{\frac{\mu \hbar \Omega_{\pm}}{2}} \left \{ \hat{A}^{\dagger}_{\pm}S_{\hat{n}}^{(\pm)} \pm \frac{1}{4 \lambda_{\pm}} \left( \hat{A}^{\dagger 3}_{\pm}H_{1,\hat{n}}^{(\pm)}+\hat{A}^{\dagger 2}_{\pm}\hat{A}_{\pm} H_{2,\hat{n}}^{(\pm)}+\hat{A}^{\dagger}_{\pm}\hat{A}_{\pm}\hat{A}^{\dagger}_{\pm}H_{3,\hat{n}}^{(\pm)}+\hat{A}_{\pm}\hat{A}^{\dagger 2}_{\pm}H_{4,\hat{n}}^{(\pm)} \right)+ \ldots \right \} + hc., \label{eq:extendp}
\end{eqnarray}}
where we have made use of the relation $\lambda_{\pm} a^{2}=\mu \Omega_{\pm}/\hbar$, and the set of diagonal operators accompanying the different powers of the deformed operators are found to be
\begin{eqnarray*}
F_{1,\hat{n}}^{(\pm)} & = & F_{\hat{n}}^{(\pm)}F_{\hat{n}+1}^{(+)}F_{\hat{n}+2}^{(\pm)}, \\
F_{2,\hat{n}}^{(\pm)} & = & G_{\hat{n}}^{(\pm)}F_{\hat{n}-1}^{(+)}F_{\hat{n}}^{(\pm)}, \label{eq:funcf2} \\
F_{3,\hat{n}}^{(\pm)} & = & F^{(\pm)2}_{\hat{n}}G_{\hat{n}+1}^{(\pm)}, \label{eq:funcf3} \\
F_{4,\hat{n}}^{(\pm)} & =& F_{\hat{n}}^{(\pm)}F_{\hat{n}+1}^{(+)}G_{\hat{n}+2}^{(\pm)}, \label{eq:funcf4}
\end{eqnarray*}
and
\begin{eqnarray*}
S_{\hat{n}}^{(\pm)} & = & \left(Q_{\hat{n}}^{(\pm)}+H_{\hat{n}+1}^{(\pm)} \right)/2, \\
H_{1,\hat{n}}^{(\pm)} & = & \left (F_{\hat{n}+2}^{(\pm)}F_{\hat{n}+1}^{(\pm)}Q_{\hat{n}}^{(\pm)}+G_{\hat{n}+1}^{(\pm)}G_{\hat{n}+2}^{(\pm)}H_{\hat{n}+3}^{(\pm)} \right)/2, \\
H_{2,\hat{n}}^{(\pm)} & = & \left(-F_{\hat{n}}^{(\pm)}F_{\hat{n}-1}^{(\pm)}H_{\hat{n}}^{(\pm)}+F_{\hat{n}-1}^{(\pm)}G_{\hat{n}}^{(\pm)}H_{\hat{n}+1}^{(\pm)} \right)/2 , \\
H_{3,\hat{n}}^{(\pm)} & =& \left( F_{\hat{n}}^{(\pm)}G_{\hat{n}+1}^{(+)}Q_{\hat{n}}^{(\pm)}+G_{\hat{n}+1}^{(\pm)}F_{\hat{n}}^{(\pm)}H_{\hat{n}+1}^{(+)} \right)/2, \\
H_{4,\hat{n}}^{(\pm)} & =& \left( G_{\hat{n}+2}^{(\pm)}F_{\hat{n}+1}^{(+)}Q_{\hat{n}}^{(\pm)}-G_{\hat{n}+1}^{(\pm)}G_{\hat{n}+2}^{(\pm)}Q_{\hat{n}+1}^{(\pm)} \right)/2.
\end{eqnarray*}
Using the commutation relation $F(\hat{n})A^{\dagger}_{\pm}=\hat{A}^{\dagger}_{\pm}F(\hat{n}+1)$, where $F(\hat{n})$ represents an arbitrary function of $\hat{n}$, together with $\hat{A}^{\dagger}_{\pm}\hat{A}_{\pm}=\hat{n}f^{2}_{\pm}(\hat{n})$ and $\hat{A}_{\pm}\hat{A}^{\dagger}_{\pm}=(\hat{n}+1)f^{2}_{\pm}(\hat{n}+1)$, (\ref{eq:extendx}) and (\ref{eq:extendp}) can be recast as follows 
\begin{eqnarray}
\hat{x}_{\pm} & \approx & \sqrt{\frac{\hbar}{2\mu \Omega_{\pm}}} \left( \hat{A}^{\dagger}_{\pm}K_{1,\hat{n}}^{(\pm)}+\hat{A}^{\dagger 3}_{\pm} K_{2,\hat{n}}^{(\pm)} + \hat{A}^{\dagger 5}_{\pm}K_{3,\hat{n}}^{(\pm)} + \ldots \right )+hc., \label{eq:extendx2} \\
\hat{p}_{\pm} & \approx & i \sqrt{\frac{\mu \hbar \Omega_{\pm}}{2}} \left( \hat{A}^{\dagger}_{\pm}J_{1,\hat{n}}^{(\pm)}+\hat{A}^{\dagger 3}_{\pm} J_{2,\hat{n}}^{(\pm)} + \hat{A}^{\dagger 5}_{\pm}J_{3,\hat{n}}^{(\pm)}+\ldots \right )+hc.,  \label{eq:extendp2}
\end{eqnarray}
where some of the diagonal operators $K_{i,\hat{n}}^{(\pm)}$ and $J_{i,\hat{n}}^{(\pm)}$ are explicitly given, to second order in $1/\lambda_{\pm}$ prefactor, by
\begin{eqnarray}
\fl K_{1,\hat{n}}^{(\pm)} & = & F_{\hat{n}}^{(\pm)}\pm \frac{1}{12 \lambda_{\pm}} \left \{ \hat{n}f^{2}_{\pm}(\hat{n})F_{2,n}^{(\pm)}+(\hat{n}+1)f^{2}_{\pm}(\hat{n}+1)F_{3,n}^{(\pm)}+(\hat{n}+2)f^{2}_{\pm}(\hat{n}+2)F_{4,n}^{(\pm)} \right \}, \label{eq:diagonalx1} \\
\fl K_{2,\hat{n}}^{(\pm)} & = & \pm \frac{1}{12 \lambda_{\pm}}F_{1,\hat{n}}^{(\pm)}+\frac{3}{160 \lambda^{2}_{\pm}} \bigg \{ \hat{n}f^{2}_{\pm}(\hat{n})F_{2,\hat{n}}^{'(\pm)}+(\hat{n}+1)f^{2}_{\pm}(\hat{n}+1) F_{3,\hat{n}}^{'(\pm)}+ (\hat{n}+2)f^{2}_{\pm}(\hat{n}+2)F_{4,n}^{'(\pm)} \nonumber \\
\fl & & +(\hat{n}+3)f^{2}_{\pm}(\hat{n}+3)F_{5,\hat{n}}^{'(\pm)}+ (\hat{n}+4)f^{2}_{\pm}(\hat{n}+4) F_{6,n}^{'(\pm)} \bigg \}, \label{eq:diagonalx2} \\
\fl K_{3,\hat{n}}^{(\pm)} & = & \frac{3}{160 \lambda^{2}_{\pm}} F_{1,\hat{n}}^{'(\pm)},
\end{eqnarray}

\begin{eqnarray}
\fl J_{1,\hat{n}}^{(\pm)} & = & S_{\hat{n}}^{(\pm)} \pm \frac{1}{4 \lambda_{\pm}} \left \{ \hat{n}f^{2}_{\pm}(\hat{n})H_{2,\hat{n}}^{(\pm)}+(\hat{n}+1)f^{2}_{\pm}(\hat{n}+1)H_{3,\hat{n}}^{(\pm)}+(\hat{n}+2)f^{2}_{\pm}(\hat{n}+2)H_{4,\hat{n}}^{(\pm)} \right \},  \\
\fl J_{2,\hat{n}}^{(\pm)}  & = & \pm \frac{1}{4 \lambda_{\pm}}H_{1,\hat{n}}^{(\pm)} + \frac{3}{32 \lambda^{2}_{\pm}} \bigg \{ \hat{n}f^{2}_{\pm}(\hat{n})H_{2,\hat{n}}^{'(\pm)} +(\hat{n}+1)f^{2}_{\pm}(\hat{n}+1)H_{3,\hat{n}}^{'(\pm)}+(\hat{n}+2)f^{2}_{\pm}(\hat{n}+2)H_{4,\hat{n}}^{'(\pm)} \nonumber \\
\fl & &+ (\hat{n}+3)f^{2}_{\pm}(\hat{n}+3)H_{5,\hat{n}}^{'(\pm)} + (\hat{n}+4)f^{2}_{\pm}(\hat{n}+4)H_{6,\hat{n}}^{'(\pm)}  \bigg \},  \\
\fl J_{3,\hat{n}}^{(\pm)} & = & \frac{3}{32 \lambda^{2}_{\pm}} H_{1,\hat{n}}^{'(\pm)},
\end{eqnarray}
and where we have introduced the notations
\begin{eqnarray*}
\fl F_{1,\hat{n}}^{'(\pm)} & = & F_{\hat{n}+4}^{(\pm)}F_{\hat{n}+3}^{(\pm)}F_{1,\hat{n}}^{(\pm)}, \qquad H_{1,\hat{n}}^{'(\pm)}  =  \left(F_{\hat{n}+3}^{(\pm)}F_{1,\hat{n}}^{(\pm)}H_{\hat{n}+5}^{(\pm)}+F_{\hat{n}+4}^{(\pm)}F_{1,\hat{n}+1}^{(\pm)}Q_{\hat{n}}^{(\pm)} \right)/ 2 ,\\
\fl F_{2,\hat{n}}^{'(\pm)} & = & F_{\hat{n}+2}^{(\pm)}F_{\hat{n}+1}^{(\pm)}F_{2,\hat{n}}^{(\pm)}, \qquad H_{2,\hat{n}}^{'(\pm)}  =  \left ( F_{\hat{n}+1}^{(+)}F_{2,\hat{n}}^{(\pm)}H_{\hat{n}+4}^{(\pm)}-F_{\hat{n}+2}^{(\pm)}F_{1,\hat{n}-1}^{(\pm)}H_{\hat{n}}^{(\pm)} \right )/2 ,\\
\fl F_{3,\hat{n}}^{'(\pm)} & = & F_{\hat{n}+2}^{(\pm)}F_{\hat{n}+1}^{(\pm)}F_{3,\hat{n}}^{(\pm)}, \qquad H_{3,\hat{n}}^{'(\pm)}  =  \left ( F_{\hat{n}+1}^{(\pm)}F_{3,\hat{n}}^{(\pm)}H_{\hat{n}+3}^{(\pm)}+F_{\hat{n}+2}^{(\pm)}F_{2,\hat{n}+1}^{(\pm)}Q_{\hat{n}}^{(\pm)} \right ) /2 ,\\
\fl F_{4,\hat{n}}^{'(\pm)} & = & F_{\hat{n}+2}^{(\pm)}F_{\hat{n}+1}^{(\pm)}F_{4,\hat{n}}^{(\pm)}, \qquad H_{4,\hat{n}}^{'(\pm)}  =  \left ( F_{\hat{n}+1}^{(\pm)}F_{4,\hat{n}}^{(\pm)}H_{\hat{n}+3}^{(\pm)}+F_{\hat{n}+2}^{(\pm)}F_{3,\hat{n}+1}^{(\pm)}Q_{\hat{n}}^{(\pm)} \right )/2 ,\\
\fl F_{5,\hat{n}}^{'(\pm)} & = & F_{\hat{n}+2}^{(\pm)}G_{\hat{n}+3}^{(\pm)}F_{1,\hat{n}}^{(\pm)}, \qquad H_{5,\hat{n}}^{'(\pm)}  =  \left ( G_{\hat{n}+3}^{(\pm)}F_{1,\hat{n}}^{(\pm)}H_{\hat{n}+3}^{(\pm)}+F_{\hat{n}+2}^{(+)}F_{4,\hat{n}+1}^{(\pm)}Q_{\hat{n}}^{(\pm)} \right ) /2, \\
\fl F_{6,\hat{n}}^{'(\pm)} & = & G_{\hat{n}+4}^{(\pm)}F_{\hat{n}+3}^{(\pm)}F_{1,\hat{n}}^{(\pm)}, \qquad H_{6,\hat{n}}^{'(\pm)}  =  \left ( -F_{\hat{n}+3}^{(\pm)}F_{1,\hat{n}}^{(\pm)}Q_{\hat{n}+3}^{(\pm)}+G_{\hat{n}+4}^{(\pm)}F_{1,\hat{n}+1}^{(\pm)}Q_{\hat{n}}^{(\pm)} \right )/2.
\end{eqnarray*}

\begin{figure}[h!] 
\begin{center}
\includegraphics[width=7cm, height=4.5cm]{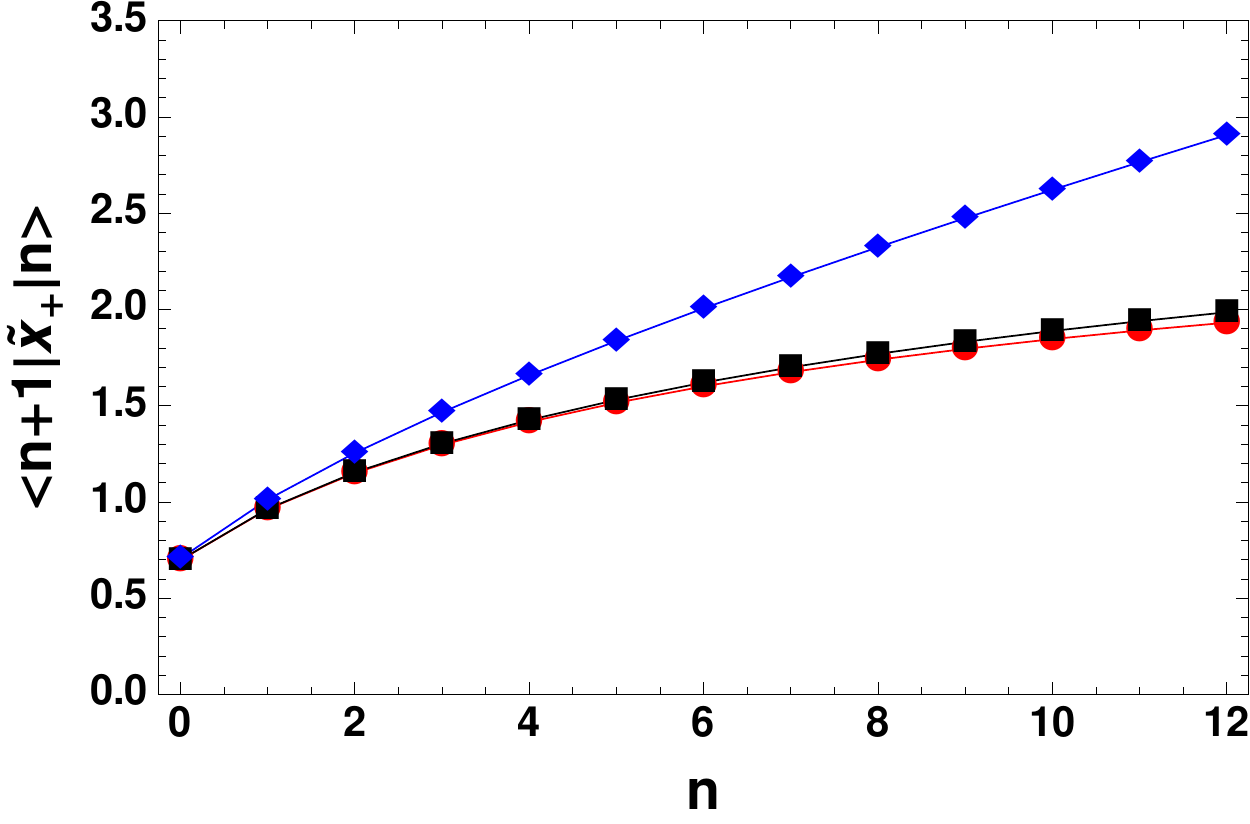} 
\includegraphics[width=7cm, height=4.5cm]{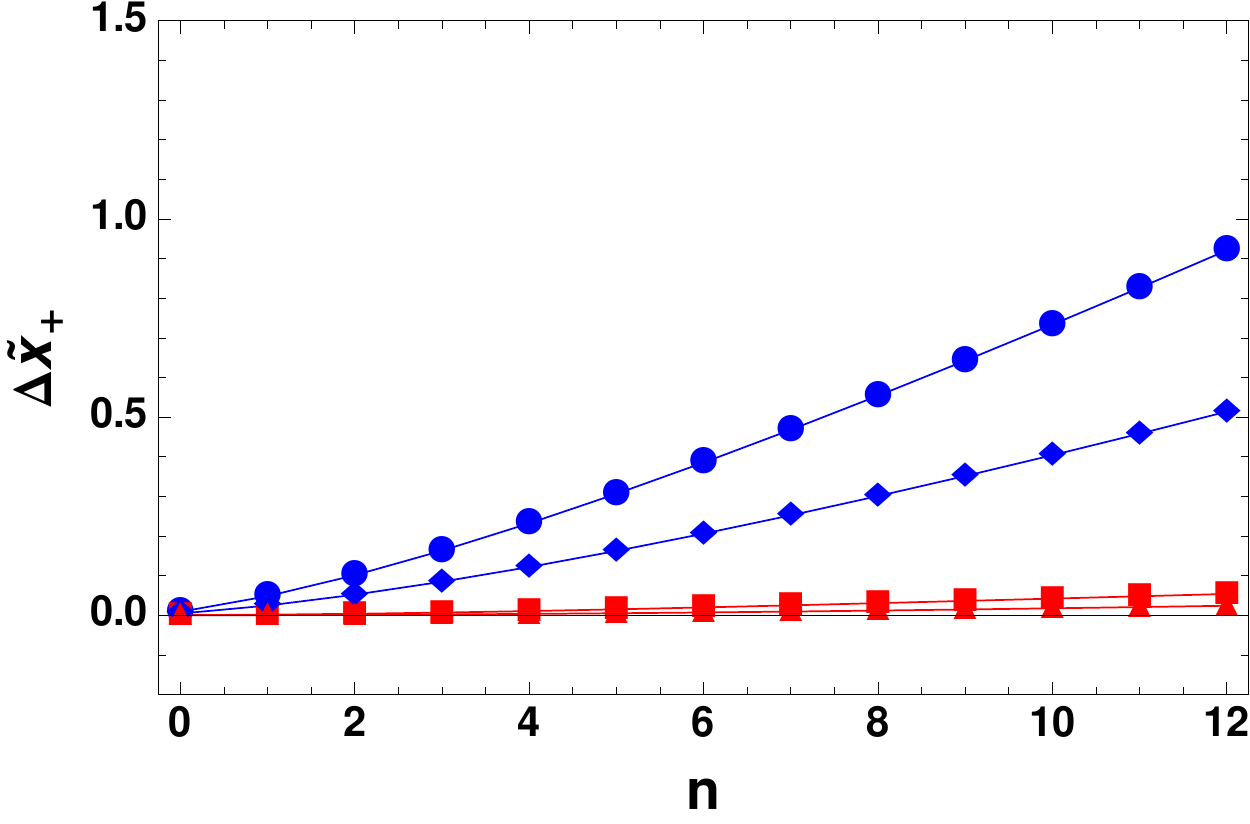} 
\caption{Left panel: Comparison of the exact matrix elements $\langle n+1 |\tilde{x}|n \rangle$ corresponding to the TPT potential with those obtained by employing the vibron (blue) and extended (red) approximations, with the latter being built up to second order in $\lambda_{+}^{-1}$, as functions of the quantum number $n$; the strength of the anharmonicity is chosen to be $\lambda_{+}^{-1}=0.05$. Right panel: Deviations $\Delta \tilde{x}_{+}=|\langle n+1 |\tilde{x}|n \rangle_{exact} - \langle n+1| \tilde{x}_{+}|n \rangle_{approx}|$ for the vibron (blue) and extended approximations (red) in considering the anharmonicity parameter values $\lambda_{+}^{-1}=0.025$ (triangle and diamond markers), $0.05$ (square and circle markers), with $a^{2}\lambda_{+}=1$. The dimensionless coordinate $\tilde{x}$ is such that $x=\sqrt{\hbar/\mu \Omega_{+}}\tilde{x}$.}
\label{fig:elematrix1}
\end{center}
\end{figure}

\begin{figure}[h!] 
\begin{center}
\includegraphics[width=7cm, height=4.5cm]{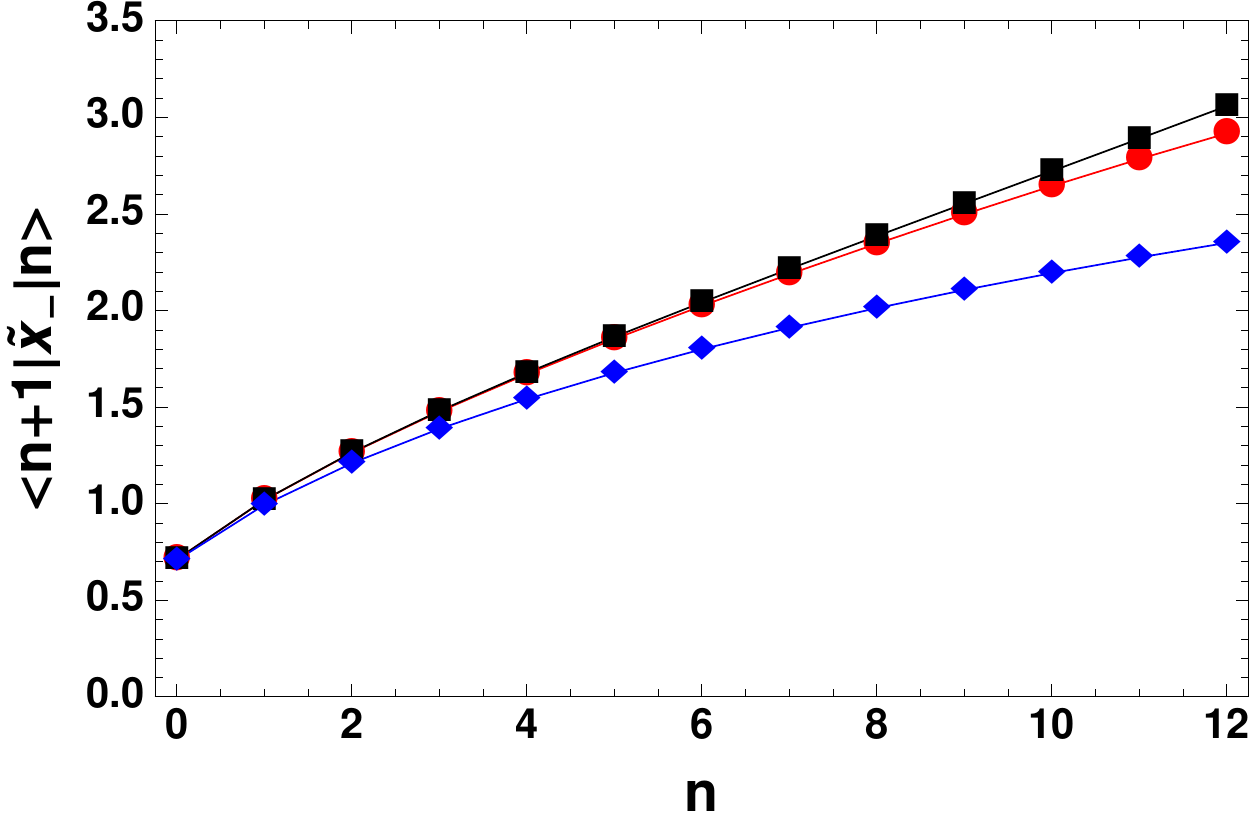} 
\includegraphics[width=7cm, height=4.5cm]{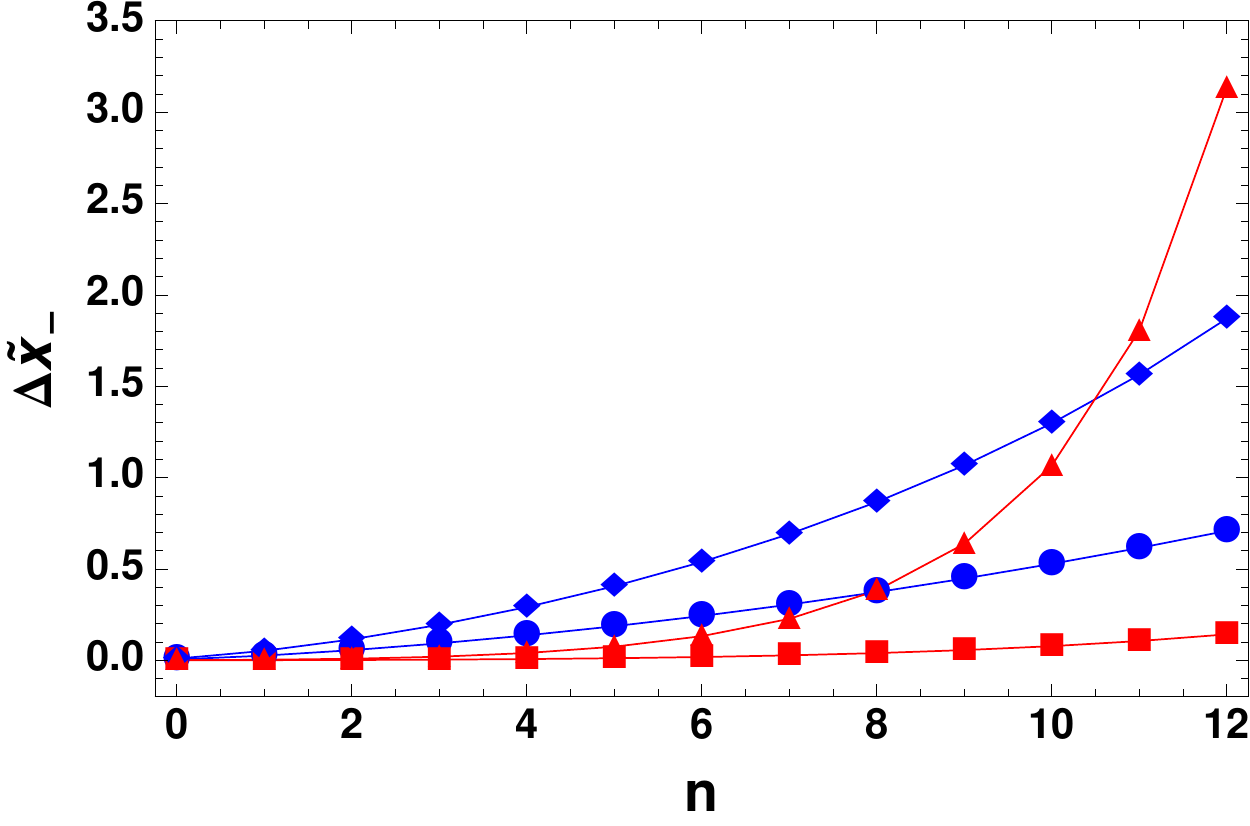} 
\caption{Left panel: Comparison of the exact matrix elements $\langle n+1 |x|n \rangle$ corresponding to the MPT potential with those obtained by employing the vibron (blue) and extended (red) approximations, with the latter being built up to first order in $\lambda_{-}^{-1}$, as functions of the quantum number $n$; the strength of the anharmonicity is chosen to be $\lambda_{-}^{-1}=0.025$. Right panel: Deviations $\Delta \tilde{x}_{-}=|\langle n+1 |x|n \rangle_{exact} - \langle n+1| x_{-}|n \rangle_{approx}|$ for the vibron (blue) and extended approximations (red) in considering the anharmonicity parameter values $\lambda_{-}^{-1}=0.025$ (square and circle markers), $0.05$ (triangle and diamond markers), with $a^{2}\lambda_{-}=1$. The dimensionless coordinate is defined as $x_{-}=\sqrt{\hbar/\mu \Omega_{+}}\tilde{x}_{-}$.}
\label{fig:elematrix2}
\end{center}
\end{figure}

In figs. \ref{fig:elematrix1} and \ref{fig:elematrix2}, a comparison of the exact matrix elements $\langle \psi_{n+1}^{(\lambda_{\pm})} |x|\psi_{n}^{(\lambda_{\pm})} \rangle$ (black line) numerically calculated from the corresponding TPT- and MPT-wave functions given by (\ref{eq:wavef1}) and (\ref{eq:wavef2}) with those obtained by means of their respective approximations (\ref{eq:extendx2}) is depicted (red line). Additionally, the outcome of the simplest position representation, commonly used in the literature, is also shown (blue line) which  consists in keeping the bare deformed operators, thereby neglecting the weighting diagonal ones whose influence, in principle, would more accurately embody the true nature of the oscillator under study. Such a representation is
\begin{eqnarray}
\hat{x}_{\pm} & \approx & \sqrt{\frac{\hbar}{2\mu \Omega_{\pm}}} (\hat{A}_{\pm}+\hat{A}^{\dagger}_{\pm}).\label{eq:simplex}
\end{eqnarray}
As in ref. \cite{lemus}, let us label (\ref{eq:extendx2}) and (\ref{eq:simplex}) as the extended and vibron  approximations, respectively. The panels on the right-hand side of the figures show the corresponding deviations from the exact results ($\Delta \tilde{x}_{\pm}=|\langle n+1 |\tilde{x}|n \rangle_{exact} - \langle n+1| \tilde{x}_{\pm}|n \rangle_{approx}|$), where $\tilde{x}$ is such that $x_{\pm}=\sqrt{\hbar/\mu_{\pm} \Omega_{\pm}}\tilde{x}_{\pm}$. Regarding the computation of the averaged position associated with the TPT potential, the  anharmonicity parameter is chosen to be $\lambda^{-1}_{+} = 0.05$. In Fig. \ref{fig:elematrix1} (left panel) one can see that the extended approximation almost stays within the line thickness of the exact result (black line) over a wide range of values of the quantum number $n$. On the contrary, the vibron approximation deviates significantly from the extended one. Deviations of this quantity from the exact result for $\lambda_{+}^{-1}=0.025$, $0.05$, and $0.1$ are plotted in the right  panel of the same figure, which reveals that even if very small values of the anharmonicity parameter $\lambda_{+}^{-1}$ are taken into account, the vibron approximation turns out to be devoid of physical usefulness in the present context. \\

For the MPT potential, the corresponding matrix elements of the coordinate $\tilde{x}_{-}$ are shown in the left panel of fig. \ref{fig:elematrix2}. We see that the vibron representation exhibits a more noticeable deviation from the exact result than that of the extended approximation for $\lambda_{-}^{-1}=0.025$, at least within the energy regime delimited in the figure. However, as the anharmonicity parameter increases, the extended approximation turns out to be totally at variance with the exact result in a moderately high-energy regime starting from $n\ge 10$, approximately, as seen in the right panel of fig. \ref{fig:elematrix2} for $\lambda_{-}^{-1}=0.025$ and $0.05$. It had already been expressly indicated in \cite{lemus}, and verified here, that this behavior is due to the very slow convergence of the series for the aforesaid energy region, and this happens even when the second-order contributions of $\lambda_{-}^{-1}$ to the series expansion of $\hat{x}_{-}$ are included; a similar situation was found to occur for the matrix elements of the momentum. Nonetheless, we note that the extended approximation may be regarded as an acceptable first-order approximation in $\lambda_{-}^{-1}$, providing we restrict our analysis to the lowest-lying states of the spectrum and for very small values of the aforesaid parameter ({\it i.e.}, $\lambda_{-}^{-1} \ll 1$). Additionally, we see that, in marked contrast to the TPT case, the vibron approximation for the MPT, albeit slightly less accurate than the extended one in this region, may provide at least a semi-quantitative description of the problem for a slightly wider range of energy eigenstates within the same parameter regime. 

\section{Van Vleck perturbation theory: general formulae for qubit-nonlinear-oscillator systems}\label{appen:b}

A very brief description of the so-called Van Vleck perturbation theory is given and pertinent general formulae are written down in this appendix following a quite similar notation as in ref. \cite{grifoni1}; see also ref. \cite{cohen} for details concerning the methodology. Firstly, let a given Hamiltonian $\hat{H}$ be decomposed as follows:
\begin{equation}
\hat{H}=\hat{H}_{0}+\hat{V}, 
\label{eq:heg}
\end{equation}
where $\hat{H}_{0}$ is the free Hamiltonian of the system at hand and $\hat{V}$ is considered to be a small perturbation proportional to a constant parameter $g$. The energy levels $E_{j,\alpha}$ of $\hat{H}_{0}$ are supposed to be  grouped in manifolds, with $\alpha$ and $j$ denoting, respectively, the index of a given manifold and $j$ the different energy levels within it. Thus,
\begin{displaymath}
\hat{H}_{0}|j,\alpha \rangle =E_{j,\alpha}|j,\alpha \rangle. 
\end{displaymath}
Secondly, provided that the parameter $g$ is small enough, which is tantamount to saying that $|\langle j,\alpha|\hat{V}|j,\beta \rangle | \ll |E_{j,\alpha}-E_{j,\beta}|$ for $\alpha \neq \beta$, as the perturbation $\hat{V}$ allows the different manifolds to be coupled, the energy eigenstates of the whole Hamiltonian (\ref{eq:heg}) are also clustered in manifolds. So, the method is aimed at constructing a transformed effective Hamiltonian $\tilde{H}_{eff} = e^{iS}\hat{H}e^{-iS}$ restricting it to act within individual manifolds in a way such that $\langle i,\alpha |\tilde{H}_{eff}|j,\beta \rangle =0$ for $\alpha \neq \beta$, preserving the same eigenvalues as $\hat{H}$. Thirdly, one then proceeds to expand $S$ and $\tilde{H}_{eff}$ in terms of the small parameter $g$ up to second order,
\begin{eqnarray}
S & = & S^{(1)}+S^{(2)}+O(g^{3}), \\
\tilde{H}_{eff} & = & \tilde{H}_{eff}^{(0)}+\tilde{H}_{eff}^{(1)}+\tilde{H}_{eff}^{(2)}+O(g^{3}),
\end{eqnarray}
for the transformation 
\begin{equation}
e^{\pm iS} = {\bf 1}\pm iS^{(1)}\pm iS^{(2)}+\frac{1}{2} iS^{(1)}iS^{(1)}+O(\bar{g}^{3})
\end{equation}
to be calculated and the effective Hamiltonian be also constructed. The perturbative constituents $S^{(\pm,1/2)}$ and $\tilde{H}_{eff}^{(\pm,1/2)}$ are calculated by making use of the fact that $\langle i,\alpha |\tilde{H}_{eff}|j,\beta \rangle =0$ for $\alpha \neq \beta$ and that $S$ has no matrix elements within the same manifold, {\it i.e.}, $\langle i,\alpha|iS^{(\pm,1/2)}|j,\alpha \rangle =0$. We now calculate the matrix elements of $S$ and $\tilde{H}_{eff}$ order by order iteratively. As to the transformation matrix $S$, one gets 
\begin{eqnarray} \fl
\langle i,\alpha |iS^{(1)}|j,\beta \rangle & = & \frac{\langle i,\alpha|V|j,\beta \rangle}{E_{i,\alpha}-E_{j,\beta}}, \quad \textrm{for} \quad \alpha \neq \beta, \\
\fl \langle i,\alpha |iS^{(2)}|j,\beta \rangle & = & - \frac{1}{2} \frac{E_{j,\beta}+E_{i,\alpha}}{E_{j,\beta}-E_{i,\alpha}} \langle i,\alpha|iS^{(1)}iS^{(1)}|j,\beta \rangle + \frac{\langle i,\alpha|iS^{(1)}\tilde{H}_{0}iS^{(1)}|j,\beta \rangle}{E_{j,\beta}-E_{i,\alpha}} \nonumber \\
& & + \frac{1}{E_{j,\beta}-E_{i,\alpha}} \left (\langle i,\alpha|ViS^{(1)}|j,\beta \rangle- \langle i,\alpha|iS^{(1)}V|j,\beta \rangle \right ).
\end{eqnarray}

The matrix elements of the the effective Hamiltonian, on the other hand, are given, up to second order in $g$, by:
{\setlength\arraycolsep{2pt}
\begin{eqnarray} \fl
\langle i,\alpha| \tilde{H}_{eff}|j, \alpha \rangle & = & \langle i,\alpha | \tilde{H}_{eff}^{(0)}|j,\alpha \rangle + \langle i,\alpha | \tilde{H}_{eff}^{(1)}|j,\alpha \rangle +\langle i,\alpha | \tilde{H}_{eff}^{(2)}|j,\alpha \rangle +O(g^{3}), \nonumber \\
& = & \langle i,\alpha | \tilde{H}_{0}|j,\alpha \rangle + \langle i,\alpha|V|j,\alpha \rangle + \frac{1}{2}  \langle i,\alpha|\left[iS^{(1)},V\right]|j,\alpha \rangle +O(g^{3}), \nonumber \\ 
& = & E_{i,\alpha} \delta_{ij}+ \langle i,\alpha|V|j,\alpha \rangle + \nonumber \\
& & +\frac{1}{2} \sum_{k,\gamma \neq \alpha} \langle i,\alpha|V|k,\gamma \rangle \langle k,\gamma|V|j,\alpha \rangle \left[\frac{1}{E_{i,\alpha}-E_{k,\gamma}}+\frac{1}{E_{j,\alpha}-E_{k,\gamma}} \right]. 
\end{eqnarray}}
So, upon applying the foregoing prescription to the problem we are interested in, that is,  
\begin{eqnarray*}
\hat{H}_{0} & = & \frac{\hbar \Delta_{Q}}{2} \tilde{\sigma}_{z}+\hbar \Omega_{\pm} \left (\hat{n}\pm \frac{\hat{n}^{2}}{2\lambda_{\pm}}\right),  \\
\hat{V} & = & -\hbar \bar{g} \left( \frac{\epsilon}{\Delta_{Q}}\tilde{\sigma}_{z} + \frac{\Delta_{0}}{\Delta_{Q}}\tilde{\sigma}_{x} \right) \left (K_{1,\hat{n}}^{(\pm)}\hat{A}_{\pm}+\hat{A}^{\dagger}_{\pm}K_{1,\hat{n}}^{(\pm)}\right),
\end{eqnarray*}
which are the constituents of our Hamiltonian model (\ref{eq:Hmodel}), one gets, for the first-order matrix elements of $S$, the following relationships:
{\setlength\arraycolsep{2pt}
\begin{eqnarray} \fl 
\langle g,m|i{S}^{(\pm,1)}|g,n\rangle & = &  \left \{ K_{1,n-1}^{(\pm)}f_{\pm}(n)\sqrt{n}\delta_{m,n-1}+K_{1,n}^{(\pm)}f_{\pm}(n+1)\sqrt{n+1}\delta_{m,n+1} \right \} \frac{\epsilon \bar{g}}{\Delta_{Q}\Omega_{m,n}^{(\pm)}} , \\ 
\fl \langle e,m|i{S}^{(\pm,1)}|e,n\rangle & = & - \left \{K_{1,n-1}^{(\pm)}f_{\pm}(n)\sqrt{n}\delta_{m,n-1}+K_{1,n}^{(\pm)}f_{\pm}(n+1)\sqrt{n+1}\delta_{m,n+1} \right \} \frac{\epsilon \bar{g}}{\Delta_{Q}\Omega_{m,n}^{(\pm)}}, \\ 
\fl \langle e,m|iS^{(\pm,1)}|g,n\rangle & = &  - \frac{\Delta_{0}}{\Delta_{Q}}\frac{K_{1,n}^{(\pm)}f_{\pm}(n+1)\sqrt{n+1}\delta_{m,n+1}}{\Delta_{Q}+\Omega_{m,n}^{(\pm)}}\bar{g},
\end{eqnarray}}
where $\Omega_{m,n}^{(\pm)} = \Omega_{\pm}(m-n)\left(1\pm\frac{m+n}{2\lambda_{\pm}}\right)$. And, for the second-order matrix elements, the general results are as follows:
{\setlength\arraycolsep{2pt}
\begin{eqnarray} \fl 
\langle g,m|iS^{(\pm,2)}|g,n\rangle & = & - \frac{K_{1,n-1}^{(\pm)}K_{1,n-2}^{(\pm)}f_{\pm}(n)f_{\pm}(n-1)\sqrt{n(n-1)}}{\Delta_{Q}^{2}\Omega_{m,n}^{(\pm)}} \\ 
& & \times \Bigg \{-\frac{\epsilon^{2}(2\Omega_{m,n-1}^{(\pm)}-\Omega_{m,n}^{(\pm)})}{2\Omega_{n-1}^{(\pm)}\Omega_{m,n-1}^{(\pm)}} +  \frac{\Delta_{0}^{2}}{\Delta_{Q}-\Omega_{m,n-1}^{(\pm)}} \Bigg \} \delta_{m,n-2} \bar{g}^{2} \nonumber \\
& - &  \frac{K_{1,n}^{(\pm)}K_{1,n+1}^{(\pm)}f_{\pm}(n+1)f_{\pm}(n+2)\sqrt{(n+1)(n+2)}}{\Delta_{Q}^{2}\Omega_{m,n}^{(\pm)}} \nonumber \\ 
& & \times \Bigg \{\frac{\epsilon^{2}(2\Omega_{m,n+1}^{(\pm)}-\Omega_{m,n}^{(\pm)})}{2\Omega_{n}^{(\pm)}\Omega_{m,n+1}^{(\pm)}} +  \frac{\Delta_{0}^{2}}{\Delta_{Q}+\Omega_{n}^{(\pm)}} \Bigg \} \delta_{m,n+2} \bar{g}^{2}, \nonumber
\end{eqnarray}}
{\setlength\arraycolsep{2pt}
\begin{eqnarray} \fl 
\langle e,m|iS^{(\pm,2)}|e,n\rangle & = &  \frac{K_{1,n-1}^{(\pm)}K_{1,n-2}^{(\pm)}f_{\pm}(n)f_{\pm}(n-1)\sqrt{n(n-1)}}{\Delta_{Q}^{2}\Omega_{m,n}^{(\pm)}} \\
& & \times \Bigg \{ \frac{\epsilon^{2}(2\Omega_{m,n-1}^{(\pm)}-\Omega_{m,n}^{(\pm)})}{2\Omega_{n-1}^{(\pm)}\Omega_{m,n-1}^{(\pm)}} + \frac{\Delta_{0}^{2}}{\Delta_{Q}+\Omega_{n-1}^{(\pm)}} \Bigg \} \delta_{m,n-2}\bar{g}^{2} \nonumber \\
& & - \frac{K_{1,n}^{(\pm)}K_{n+1}^{(\pm)}f_{\pm}(n+1)f_{\pm}(n+2)\sqrt{(n+1)(n+2)}}{\Delta_{Q}^{2}\Omega_{m,n}^{(\pm)}} \nonumber \\ 
& & \times \Bigg \{ \frac{\epsilon^{2}(2\Omega_{m,n+1}^{(\pm)}-\Omega_{m,n}^{(\pm)})}{2\Omega_{n}^{(\pm)}\Omega_{m,n+1}^{(\pm)}} - \frac{\Delta_{0}^{2}}{\Delta_{Q}+\Omega_{m,n+1}^{(\pm)}} \Bigg \} \delta_{m,n+2}\bar{g}^{2}, \nonumber
\end{eqnarray}}
{\setlength\arraycolsep{2pt}
\begin{eqnarray}
\fl \langle e,m|iS^{(\pm, 2)}|g,n\rangle & = & -\frac{\epsilon \Delta_{0}}{ \Delta_{Q}^{2}} \frac{K_{1,n}^{(\pm)2}f^{2}_{\pm}(n+1)(n+1)}{\Delta_{Q}+\Omega_{m,n}^{(\pm)}}  \\ 
& & \times \left \{ \frac{1}{2} \frac{\Delta_{Q}-\Omega_{m,n}^{(\pm)}}{\Omega_{m,n+1}^{(\pm)}(\Delta_{Q}+\Omega_{n}^{(\pm)})} - \frac{\Delta_{Q}}{\Delta_{Q}+\Omega_{n}^{(\pm)}} \left( \frac{1}{\Omega_{n}^{(\pm)}}+\frac{1}{\Omega_{m,n+1}^{(\pm)}} \right) \right \} \delta_{m,n} \bar{g}^{2} \nonumber \\
& & + \frac{\epsilon \Delta_{0}}{ \Delta_{Q}^{2}} \frac{K_{1,n-1}^{(\pm)2}f^{2}_{\pm}(n)n}{\Delta_{Q}+\Omega_{m,n}^{(\pm)}} \left \{ \frac{1}{2}\frac{\Delta_{Q}+\Omega_{m,n}^{(\pm)}}{\Omega_{n-1}^{(\pm)}(\Delta_{Q}+\Omega_{m,n-1}^{(\pm)})}- \frac{1}{\Omega_{n-1}^{(\pm)}}+ \frac{1}{\Omega_{m,n-1}^{(\pm)}} \right \} \delta_{m,n}\bar{g}^{2} \nonumber \\
& & - \frac{\epsilon \Delta_{0}}{\Delta_{Q}^{2}} \frac{K_{1,n}^{(\pm)}K_{1,n+1}^{(\pm)}f_{\pm}(n+1)f_{\pm}(n+2)\sqrt{(n+1)(n+2)}}{\Delta_{Q}+\Omega_{m,n}^{(\pm)}} \nonumber \\ 
& & \Bigg \{ \frac{1}{2} \frac{\Delta_{Q}-\Omega_{m,n}^{(\pm)}}{\Omega_{m,n+1}^{(\pm)}(\Delta_{Q}+\Omega_{n}^{(\pm)})} +\frac{1}{2} \frac{\Delta_{Q}+\Omega_{m,n}^{(\pm)}}{\Omega_{n}^{(\pm)}(\Delta_{Q}+\Omega_{m,n+1}^{(\pm)})} \nonumber \\
& & - \frac{\Delta_{Q}}{\Delta_{Q}+\Omega_{n}^{(\pm)}} \left( \frac{1}{\Omega_{m,n+1}^{(\pm)}}+\frac{1}{\Omega_{n}^{(\pm)}} \right) \Bigg \} \delta_{m,n+2} \bar{g}^{2} \nonumber \\
& & - \frac{\epsilon \Delta_{0}}{\Delta_{Q}^{2}} \frac{K_{1,n-1}^{(\pm)}K_{1,n-2}^{(\pm)}f_{\pm}(n)f_{\pm}(n-1)\sqrt{n(n-1)}}{\Delta_{Q}+\Omega_{m,n}^{(\pm)}} \left \{\frac{\Omega_{m,n-1}^{(\pm)}-\Omega_{n-1}^{(\pm)}}{\Omega_{n-1}^{(\pm)}\Omega_{m,n-1}^{(\pm)}} \right \} \delta_{m,n-2}\bar{g}^{2}. \nonumber
\end{eqnarray}}
{\setlength\arraycolsep{2pt}
\begin{eqnarray}
\fl \langle g,m|i{S}^{(\pm,1)}i{S}^{(\pm,1)}|g,n\rangle & = & \langle e,m|i{S}^{(\pm,1)}i{S}^{(\pm,1)}|e,n\rangle,  \\ 
 & = & \frac{\epsilon^{2}}{\Delta_{Q}^{2}} \Bigg \{ -\frac{K_{1,n-2}^{(\pm)}K_{1,n-1}^{(\pm)}f_{\pm}(n-1)f_{\pm}(n)\sqrt{n(n-1)}}{\Omega_{m,n-1}^{(\pm)}\Omega_{n-1}^{(\pm)}} \delta_{m,n-2}\nonumber \\
 & & +\frac{K_{1,n+1}^{(\pm)}K_{1,n}^{(\pm)}f_{\pm}(n+1)f_{\pm}(n+2)\sqrt{(n+1)(n+2)}}{\Omega_{m,n+1}^{(\pm)}\Omega_{n}^{(\pm)}}\delta_{m,n+2} \Bigg \}\bar{g}^{2}, \nonumber \\
\fl \langle e,m|i{S}^{(\pm,1)}i{S}^{(\pm,1)}|g,n\rangle & = & \frac{\epsilon \Delta_{0}}{\Delta_{Q}^{2}} \Bigg \{ \frac{K_{1,n}^{(\pm) 2}f^{2}_{\pm}(n+1)(n+1)}{\Omega_{m,n+1}^{(\pm)}(\Delta_{Q}+\Omega_{n}^{(\pm)})} + \frac{K_{1,n-1}^{(\pm) 2}f^{2}_{\pm}(n)n}{\Omega_{n-1}^{(\pm)}(\Delta_{Q}+\Omega_{m,n-1}^{(\pm)})} \Bigg \} \delta_{m,n} \bar{g}^{2} \nonumber \\
 &  & + \frac{\epsilon \Delta_{0}}{\Delta_{Q}} K_{1,n}^{(\pm)}K_{1,n+1}^{(\pm)}f_{\pm}(n+1)f_{\pm}(n+2)\sqrt{(n+1)(n+2)}  \nonumber \\ 
  & & \quad \times  \frac{\Omega_{n}^{(\pm)}-\Omega_{m,n+1}^{(\pm)}}{\Omega_{m,n+1}^{(\pm)}\Omega_{n}^{(\pm)}(\Delta_{Q}+\Omega_{n}^{(\pm)})(\Delta_{Q}+\Omega_{m,n+1}^{(\pm)})}  \delta_{m,n+2} \bar{g}^{2}. 
\end{eqnarray}}
This set of generalized formulae reproduces the corresponding results obtained for the qubit-linear-oscillator case \cite{grifoni1} when $f_{\pm}(\hat{n}) =1$ (or, equivalently, in the limit $\lambda_{\pm}^{-1} \to 0$).

\end{document}